\documentclass[final]{IEEEtran}
\usepackage{amsmath,amssymb,amsfonts,mathrsfs,bm,amsthm}
\usepackage{amstext}
\usepackage{upgreek}

\usepackage{setspace}
\usepackage{multicol}
\usepackage{indentfirst}
\usepackage{graphicx}
\usepackage{paralist}
\usepackage{multirow}
\usepackage{tabularx}
\usepackage[noadjust]{cite}

\usepackage{booktabs}
\usepackage{subfigure}
\usepackage{color}
\usepackage{textcomp}

\IEEEoverridecommandlockouts
\allowdisplaybreaks
\usepackage{algorithm,algorithmicx}
\floatname{algorithm}{\bf Algorithm}

\usepackage[noend]{algpseudocode}
\algrenewcommand{\algorithmiccomment}[1]{\hfill // #1}

\newcommand\Algphase[1]{%
	\vspace*{0.5\baselineskip}\Statex\hspace*{-\algorithmicindent}\texttt{#1}%
}

\newcommand\Algtop[1]{%
	\vspace*{0\baselineskip}\Statex\hspace*{-\algorithmicindent}\texttt{#1}%
}

\makeatletter
\newcommand{\algmargin}{\the\ALG@thistlm}
\makeatother
\algnewcommand{\parState}[1]{\State 
	\parbox[t]{\dimexpr\linewidth-\algmargin}{\strut #1\strut}}

\newtheorem{definition}{\bf Definition}
\newtheorem{proposition}{\bf Proposition}

\newtheorem{lemma}{\bf Lemma}

\usepackage{accents}

\allowdisplaybreaks

\definecolor{blue}{rgb}{0,0.24,0.54}
\begin{document}
\title{Deep Learning Aided Packet Routing in Aeronautical \emph{Ad-Hoc} Networks Relying on Real Flight Data: From Single-Objective to Near-Pareto Multi-Objective Optimization} %
\author{Dong Liu, Jiankang Zhang, Jingjing Cui, Soon-Xin Ng, Robert G. Maunder, and Lajos Hanzo
	
\thanks{D. Liu is with the School of Cyber Science and Technology, Beihang University, Beijing 100191, China (email: dliu@buaa.edu.cn).  }
\thanks{J. Cui, S. X. Ng, R, G. Maunder, and L. Hanzo are with the School of Electronics and Computer Science, the University of Southampton, Southampton SO17 1BJ, UK (email: jingj.cui@soton.ac.uk; sxn@ecs.soton.ac.uk; rm@ecs.soton.ac.uk; lh@ecs.soton.ac.uk).}
\thanks{J. Zhang is with the Department of Computing and Informatics, Bournemouth University, Bournemouth BH12 5BB, U.K. (e-mail: jzhang3@bournemouth.ac.uk).}
\thanks{L. Hanzo would like to acknowledge the financial support of the Engineering and Physical Sciences Research Council projects EP/P034284/1 and EP/P003990/1 (COALESCE) as well as of the European Research Council's Advanced Fellow Grant QuantCom (Grant No. 789028)}
}

\maketitle
\begin{abstract}
Data packet routing in aeronautical \emph{ad-hoc} networks (AANETs) is challenging due to their high-dynamic topology. In this paper, we invoke deep learning (DL) to assist routing in AANETs. We set out from the single objective of minimizing the end-to-end (E2E) delay. Specifically, a deep neural network (DNN) is conceived for mapping the local geographic information observed by the forwarding node into the information required for determining the optimal next hop. The DNN is trained by exploiting the regular mobility pattern of commercial passenger airplanes from historical flight data. After training, the DNN is stored by each airplane for assisting their routing decisions during flight  relying solely on local geographic information. Furthermore, we extend the DL-aided routing algorithm to a multi-objective scenario, where we aim for simultaneously minimizing the delay, maximizing the path capacity and maximizing the path lifetime. Our simulation results based on real flight data show that the proposed DL-aided routing outperforms existing position-based routing protocols in terms of its E2E delay, path capacity as well as path lifetime, and it is capable of approaching the Pareto front that is obtained using global link information.  
\end{abstract}


\vspace{-2mm}
\section{Introduction}
Next-generation wireless networks are envisaged to support truly global communications, anywhere and anytime~\cite{huang2019airplane}. Current in-flight Internet access supported by geostationary satellites or direct air-to-ground (A2G) communications typically exhibit either high latency or limited coverage. Aeronautical \emph{ad-hoc} networks (AANETs) are potentially capable of extending the coverage of A2G networks by relying on commercial passenger airplanes to act as relays for forming a self-configured wireless network via multihop air-to-air (A2A) communication links~\cite{zhang2019aeronautical,luo2018aeromrp}.

Due to the high velocity of aircraft and the distributed nature of \emph{ad-hoc} networking, one of the fundamental challenges in AANETs is to design an efficient routing protocol for constructing an appropriate routing path at any time~\cite{bujari2018would}. Traditional topology-based \emph{ad-hoc} routing protocols~\cite{li2007routing}, such as the \emph{ad-hoc} on-demand distance vector (AODV)~\cite{aodv} based solutions, usually require each node to locally store a routing table specifying the next hop. The routing table, however, has to be refreshed whenever the network topology changes during a communication session, hence imposing excessive signaling overhead and latency in AANETs. Although substantial research efforts have been invested in improving the stability of routing in AANETs~\cite{sakhaee2006global,luo2017multiple}, they have a limited ability to update their routing tables for prompt adaptation in high-dynamic scenarios. Reinforcement learning based routing algorithms, such as Q-routing~\cite{boyan1994packet} and its variants~\cite{mammeri2019reinforcement} were proposed for improving the adaptability of routing in dynamic environments. However, they require frequent information exchange for trail-and-error in order to update the Q-table (acting as the routing table) in an online manner, which still suffer from the above issues when the network topology is highly dynamic. 

\begin{table*}
	\centering
	\footnotesize
	\setlength{\extrarowheight}{1.5pt}
	\caption{Comparison with existing routing algorithms for AANETs}
	\begin{tabular}{|m{3.7cm}|m{1.4cm}|m{1.3cm}|m{2cm}|m{1.3cm}|m{1.2cm}|m{1.2cm}|m{1.2cm}|}
		\hline
		Routing Algorithms & \parbox{1.5cm}{ AODV~\cite{aodv} \\ MUDOR~\cite{sakhaee2006global}} & GPSR~\cite{karp2000gpsr} GLSR~\cite{medina2011geographic} A-GR~\cite{wang2013gr} & \parbox{2.2cm}{Q-Routing~\cite{boyan1994packet} \&\\ its~variants~\cite{mammeri2019reinforcement}} & MQSPR~\cite{luo2017multiple} & MOEA~\cite{yetgin2012multi}  & Proposed POMOR & Proposed DL-aided Routing \\
		\hline
		\parbox{3.8cm}{{\vspace*{1mm}\bf Centralized:}\\ Global Topology Information\\
			Low-Dynamic Scenario \vspace{1mm}} &       &       &       &       & \parbox{1.2cm}{\centering \vspace{1.5em}$\surd$}     & \parbox{1.2cm}{\centering \vspace{1.5em}$\surd$}      &  \\ \hline
		\parbox{3.8cm}{{\vspace*{1mm} \bf Ad-Hoc (Topology-Based):} \\Routing Table  \\ Low-Dynamic Scenario \vspace{1mm}} & \parbox{1.5cm}{\centering \vspace{1.5em}$\surd$}      &       & \parbox{2cm}{\centering \vspace{1.5em}$\surd$}     & \parbox{1.5cm}{\centering \vspace{1.5em}$\surd$}     &       &       &  \\\hline
		\parbox{3.8cm}{\vspace{1mm}{\bf Ad-Hoc (Position-Based):} \\ Local Geographic Information \\ High-Dynamic Scenario \vspace{1mm}} &       & \parbox{1.3cm}{\centering \vspace{1.5em}$\surd$}     &       &       &       &       & \parbox{1.2cm}{\centering \vspace{1.5em}$\surd$} \\\hline
		\bf Single Objective & \parbox{1.5cm}{\centering $\surd$}     & \parbox{1.3cm}{\centering $\surd$}     & \parbox{2cm}{\centering $\surd$}     &       &       &       & \parbox{1.2cm}{\centering $\surd$} \\\hline
		\bf Mutliple Objective &       &       &       & \parbox{1.2cm}{\centering $\surd$}     & \parbox{1.2cm}{\centering $\surd$}     & \parbox{1.2cm}{\centering $\surd$}     & \parbox{1.2cm}{\centering $\surd$} \\\hline
		\bf Find All Pareto-Optimal Paths &       &       &       &        &       & \parbox{1.2cm}{\centering $\surd$}     &  \\
		\hline
	\end{tabular}%
	\label{tab:gap}%
\end{table*}%

By contrast, another family of \emph{ad-hoc} routing protocols, namely position-based (or geographic) routing~\cite{mauve2001survey}, only requires the position information of the single-hop neighbors and of the destination for determining the next hop. Since it does not have to maintain routing tables, geographic routing finds new routes almost instantly, when the topology changes. Because the  position information required for determining the next hop can be readily obtained by each airplane using the automatic dependent surveillance-broadcast (ADS-B) system on board, geographic routing is more appealing in AANETs. Greedy perimeter stateless routing (GPSR)~\cite{karp2000gpsr} was one of the most popular geographic routing protocols, which has also inspired various extensions~\cite{medina2011geographic,wang2013gr} in AANETs. The core idea of greedy routing is to forward the packet to the specific neighbor that is geographically closest to the destination. In \cite{medina2011geographic}, greedy routing was improved for avoiding congestion by considering the buffer queue status of the next-hop candidates. In \cite{wang2013gr}, the mobility information was further taken into consideration for choosing a more stable next hop. However, greedy routing~\cite{karp2000gpsr,medina2011geographic,wang2013gr} is stifled when no neighbor is closer to the destination than the forwarding node (this situation is term as the \emph{communication void}). 

To elaborate, the limitation of position-based routing arises from the fact that the nodes are unaware of the entire network topology. Therefore, one of our ambitious goals is to \emph{enable
the forwarding node to infer the information required for avoiding the communication void issue from its local observation}. Although the topology of AANETs evolves dynamically, it exhibits certain patterns, since both the flight trajectories and takeoff times are pre-planned and normally repeated on a weekly basis. This suggests that the local geographic information may be strongly correlated with that of the whole topology, hence this correlation may be exploited using historical flight data.

Another typical limitation of the existing routing protocols designed for AANETs is that only a single-component objective function (OF) is optimized for improving a specific network performance metric. However, the overall network performance should be characterized by multiple metrics, such as the end-to-end (E2E) packet delay, path capacity and path lifetime. Accordingly, multi-objective optimization (MOO)~\cite{chankong1983multiobjective} can be adopted, where all components of the OF are optimized simultaneously. Since multiple objectives typically conflict with each other, they may not achieve their respective optima at the same time. Hence, in contrast to optimizing a single-component OF, typically there is no single globally optimal solution in MOO, which is the best with respect to all objectives. By contrast, the ultimate goal of MOO is to discover all the \emph{Pareto-optimal} solutions that constitute the entire \emph{Pareto front}, where none of the objective can be improved without scarifying at least one of the other objectives~\cite{chankong1983multiobjective}. To expound a little further, the Pareto front characterizes the optimal tradeoff relationship among multiple potentially conflicting objectives. As a benefit, the network operator may opt for any of the Pareto-optimal solutions along the Pareto front according to the requirements of different services. For example, for file downloading, it is more important to increase the capacity than reducing the E2E delay, while for voice calls, it is more important to reduce the E2E delay.

MOO has been leveraged in diverse problems in wireless networks~\cite{fei2016survey,wang2020thirty} and also been used for routing in AANET~\cite{luo2017multiple}. However,  multiple performance metrics were combined by a linear weighted sum in~\cite{luo2017multiple}, by which some Pareto-optimal solutions may not be found, when the Pareto front is non-convex. Alternatively, bio-inspired metaheuristics, such as multi-objective evolutionary algorithms (MOEAs), can be leveraged for solving the multi-objective routing problem~\cite{yetgin2012multi},  which have the potential to find the Pareto-optimal solutions even in non-convex scenarios. However, MOEAs require global knowledge regarding the status (e.g., delay, capacity, lifetime) of every single possible link in the network for running the optimization, which is not feasible in large-scale AANETs.

\begin{table*}
	\vspace{-5mm}
	\caption{Summary of Main Notations}
	\label{tab:notation}
	\scriptsize
	\centering
	\setlength{\extrarowheight}{0.5pt}
	\vspace{-2mm}
	\begin{tabular}{|l|l||l|l|}\hline
		$\mathcal{N}^{[t]}$ & Set of nodes available at TS $t$ & $p_n$ & The $n$th node on path $\bm p$ \\ \hline
		${\bm s}_i^{[t]}, \bar {\bm s}_i^{[t]}$ & Spherical/Cartesian coordinates of node $i$ at TS $t$ & $D^{[t]}(\bm p)$ & Delay of path $\bm p$  at TS $t$\\ \hline
		$\theta_i^{[t]}$ &  Latitude of node $i$ at TS $t$ & $C^{[t]}(\bm p)$ & Capacity of path $\bm p$ at TS $t$\\ \hline
		$\varphi_i^{[t]}$ &  Longitude of node $i$ at TS $t$ & 	$L^{[t]}(\bm p)$ & Lifetime of path $\bm p$ at TS $t$\\ \hline
		$h_i^{[t]}$ &  Altitude of node $i$ at TS $t$ &  $L^{[t]}(i, j)$ & Lifetime of link $i\to j$ starting from TS $t$ \\ \hline		
		$R_{\sf e}$ & Earth radius & 		$\bm p^*_{i_{\sf s}}$ & Optimal path from node $i_{\sf s}$ to $i_{\sf d}$\\ \hline
		$d^{[t]}(i, j)$ & Distance between nodes $i$ and $j$ at TS $t$ & 		$D_*^{[t]}(j, i_d)$ & Minimum delay from $j$ to $i_{\sf}$ at TS $t$ \\\hline
		$d_{\sf th}(h_i^{[t]}, h_j^{[t]})$ & Maximum distance of a direct link & 		$\hat D_*^{[t]}(j, i_d)$ & Estimated minimum delay using DNN\\\hline
		$\mathcal{B}_i^{[t]}$ & Neighbor set of node $i$ at TS $t$ & 		$\tilde D_*^{[t]}(j, i_d)$ & Estimated minimum delay using feedback\\\hline
		$b_k^{[t]}(i)$ & The $k$th neighbor of node $i$ at TS $t$ & 		$\bm x_i^{[t]}$ & Local information observed by node $i$ at TS $t$ \\\hline
		$C^{[t]}(i,j)$ & The capacity of link $i\to j$ at TS $t$ & 		$\bm y_i^{[t]}, \hat {\bm y}_i^{[t]}$ & Desired/Actual output of the DNN at node $i$ \\\hline
		$W, f$ & Transmission bandwidth/Carrier frequency & 		$\bm f(\cdot; \bm \theta)$ & DNN with parameter $\bm \theta$ \\\hline
		$P, \sigma^2$ & Transmit/noise power & 		$\mathcal C_i^{[t]}$ & Next-hop candidate set of node $i$ \\\hline
		$G_{\sf t}, G_{\sf r} $ & Transmit/Receive antenna gain  & 		$\mathcal M_i^{[t]}$ & Mutual candidate set of node $i$, see \eqref{eqn:M} \\\hline
		$D^{[t]}(i,j)$ & Total delay of link $i \to j$ 	&	$m_k$ & The $k$th element in the sorted mutual candidate set \\\hline
		$D^{[t]}_{\sf que}(i)$ & Queuing delay at node $i$ & 		$\mathcal {\bar C}_i^{[t]}$ & Defined in \eqref{eqn:barC}\\\hline
		$D^{[t]}_{\sf tr}(i, j)$ & Transmission delay of link $i \to j$  & $X_*^{[t]}(j, i_{\sf d}, \varepsilon_C, \varepsilon_L)$ &  Defined in \eqref{eqn:next-hop-multi} \\\hline
		$D^{[t]}_{\sf pr}(i, j)$ & Propagation delay of link $i \to j$ & $\hat X (j, i_{\sf d}, \varepsilon_C, \varepsilon_L)$ &  The estimate of $X_*^{[t]}(j, i_{\sf d}, \varepsilon_C, \varepsilon_L)$\\\hline
		$S$ & Packet size & $\tilde X (j, i_{\sf d}, \varepsilon_C, \varepsilon_L)$ &  The estimate of $X_*^{[t]}(j, i_{\sf d}, \varepsilon_C, \varepsilon_L)$ using feedback\\\hline
	\end{tabular}
\end{table*}
Therefore, another radical goal of this paper is to \emph{devise routing algorithm simultaneously optimizing multiple performance metrics based on local information in a distributed manner}. As we mentioned before, the local geographic information may be strongly correlated with the global topology due to the regular mobility pattern of commercial passenger airplanes. In this context, deep neural networks (DNNs)~\cite{lecun2015deep} are capable of learning a direct mapping from the input features to the desired output.

Against this background, we conceive a bespoke DL technique for learning routing in AANETs. We commence with a single objective aimed at minimizing the E2E delay and then extend our investigations to a multi-objective scenario, where the E2E delay, path capacity, and path lifetime are jointly considered as components of the OF. The contributions of this paper are summarized as follows:
\begin{enumerate}
	\item We propose a  DL-aided routing algorithm for minimizing the E2E delay in AANETs. Specifically, we use a DNN that takes the coordinates of the next-hop candidates as its input and outputs the information required for determining the optimal next hop. The DNN is trained offline by supervised learning, where the training labels are obtained by running the shortest path algorithm using historical flight data. With the aid of the pre-trained DNN, the optimal next hop can be determined promptly, solely based on local geographic information. 
	\item We further extend our DL-aided routing algorithm for handling multi-component OF, namely simultaneously minimizing the delay, maximizing the path capacity and maximizing the path lifetime. To generate training labels for the DNN, we propose a Pareto-optimal multi-objective routing (POMOR) algorithm using the global link information to obtain all the Pareto-optimal solutions at a moderate polynomial complexity.
	\item Our simulation results based on real flight data collected over the North Atlantic Ocean and the European Continent show that:
	\begin{itemize}
		\item For minimum-delay routing, our DL-aided routing outperforms the existing geographic routing and it is capable of approaching the performance of the optimal routing that relies on global link information.
		\item For multi-objective routing, our proposed POMOR algorithm using global link information can find the Pareto front for visualizing and analyzing the optimal tradeoff between multiple performance metrics. When  relying solely on local information, our DL-aided routing is capable of discovering paths that achieve a performance closely the Pareto front obtained by POMOR.
	\end{itemize}
\end{enumerate}

In Table~\ref{tab:gap}, we boldly and explicitly contrast our work to the most pertinent routing algorithms designed for AANETs. The rest of the paper is organized as follows. In Section~II, we introduce the system model. In Section~III, we propose our DL-aided routing algorithm for single objective optimization minimizing the E2E delay. In Section~IV, we extend the DL-aided routing to a challenging multi-objective scenario. Our simulations results are provided in Section~V, and finally, Section~VI concludes the paper.

\section{System Model}

Consider an AANET formed by multiple passenger airplanes and a ground station (GS). Let $i$ denote the node's index and $\mathcal{N}^{[t]}$ denote the set of node indices at timestamp (TS) $t$. The position of node $i$ at TS $t$ is denoted by a 3D-vector $\bm s^{[t]}_i = \big[\theta_i^{[t]}, \varphi_i^{[t]}, h_i^{[t]}\big]$
where $\theta_i^{[t]}$, $\varphi_i^{[t]}$, and $h_i^{[t]}$ denote the latitude, longitude, and altitude of node $i$ at TS $t$, respectively.  The
main notations to be used throughout the article are summarized
in Table.~\ref{tab:notation}.

In Cartesian coordinates, the position of node $i$ can be converted as
\begin{align}
\bar{\bm s}_i^{[t]} = \Big[&(R_{\sf e} + h_i^{[t]}) \cos\theta_i^{[t]}\cos\varphi_i^{[t]}, (R_{\sf e} + h_i^{[t]}) \cos\theta_i^{[t]}\sin\varphi_i^{[t]}, \nonumber\\
& (R_{\sf e} + h_i^{[t]}) \sin\varphi_i^{[t]}\Big]
\end{align}
where $R_{\sf e} = 6371$ km is the earth radius. Then, the Euclidean distance between node $i$ and node $j$ at TS $t$ can be expressed as $d^{[t]}(i,j) = \big\Vert \bar{\bm s}_i^{[t]} - \bar{\bm s}_j^{[t]}\big\Vert$.

\begin{figure}[!htb]
	\centering
	\includegraphics[width=0.3\textwidth]{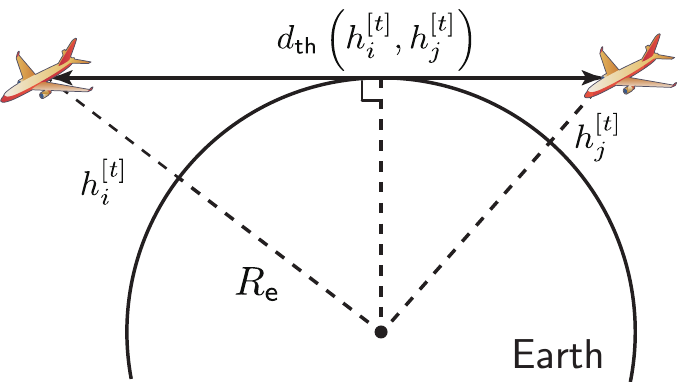}
	\caption{The maximum distance for direct communication between two nodes above the horizon.}
	\label{fig:dth}
\end{figure}
 
A pair of nodes can establish direct communications when they have direct visibility, i.e., when they are both above the horizon, as shown in Fig.~\ref{fig:dth}. Then, the maximum communication distance between nodes $i$ and $j$ is given by 
\begin{equation}
d_{\sf th}\left(h_i^{[t]}, h_j^{[t]}\right)= \sqrt{(h_i^{[t]} + R_{\sf e})^2 - R_{\sf e}^2} + \sqrt{(h_j^{[t]} + R_{\sf e})^2 - R_{\sf e}^2}.
\end{equation}
Therefore, the condition that nodes $i$ and $j$ can establish a direct communication link is $d^{[t]}(i,j) \leq d_{\sf th}(h_i^{[t]}, h_j^{[t]})$.

Let $\mathcal{B}^{[t]}_i = \{j |  d^{[t]}(i,j) \leq d_{\sf th}(h_i^{[t]}, h_j^{[t]})\}$ denote the set of nodes that can connect to node $i$, i.e., the \emph{neighbors} of node~$i$. Moreover, we rank the neighbors by their distance to the destination in ascending order and let $b_{k}^{[t]}(i)$ denote the $k$th neighbor of node $i$ at TS~$t$.

Upon assuming free-space path loss above the clouds, the capacity of the direct link from node $i$ to node $j$ can be expressed as
\begin{equation}
C^{[t]}(i,j) = W\log_2 \left[ 1 + \frac{PG_{\sf t}G_{\sf r}}{\sigma^2} \left(\frac{c}{4\pi f d^{[t]}(i,j)}\right)^2 \right],
\end{equation}
where $W$ is the transmission bandwidth, $P$ is the transmit power, $G_{\sf t}$ and $G_{\sf r}$ denote the transmit and receive antenna gains, respectively, $\sigma^2$ is the noise power, $c=3\times 10^8$ m/s is the speed of light, and $f$ is the carrier frequency.

Let $S$ denote the packet size. The delay of sending a packet from node~$i$ to node $j$ is composed by 
\begin{equation}
D^{[t]}(i,j) = D_{\sf que}^{[t]}(i) +D_{\sf tr}^{[t]}(i, j) + D_{\sf pr}^{[t]}(i,j), \label{eqn:Dlink}
\end{equation}
where $D_{\sf que}^{[t]}(i)$ denotes the duration that the packet spent in the buffer queue of node $i$, which can be formulated as $D_{\sf que}^{[t]}(i) = \sum_{q=1}^{Q_i^{[t]}} \frac{S}{C^{[t]}(i, n_{q})}$ with $Q_i^{[t]}$ denoting the number of packets in the buffer queue of node $i$ and $n_q$ denoting the next hop for the $q$th packet in the queue, $D_{\sf tr}^{[t]}(i,j) = \frac{S}{C^{[t]}(i,j)}$ is the transmission delay, and $D_{\sf pr}^{[t]}(i,j) = \frac{d^{[t]}(i,j)}{c}$ is the propagation delay from node $i$ to node $j$.

Since the distance between a source node $i_{\sf s}$ and a destination node $i_{\sf d}$ may exceed the direct communication range, the packet may traverse through multiple nodes until finally reaching $i_{\sf d}$. Let $\bm p \triangleq (p_1, p_2, \cdots, p_{N})$ denote a path having $N-1$ hops connecting nodes $i_{\sf s}$ and $i_{\sf d}$, where $p_n$ denotes the index of the $n$th node on path $\bm p$. In particular, we have $p_1 = i_{\sf s}$ and $p_{N} = i_{\sf d}$. 

We consider three metrics for routing path selection. The first one is the E2E delay of the path, which is the summation over the delay of each link along the path and can be formulated as
\begin{equation}
D^{[t]}(\bm p) = \sum_{n=1}^{N-1} D^{[t]}(p_n, p_{n+1}).
\end{equation}

The second one is the path capacity, which is limited by the lowest link capacity along the path as
\begin{equation}
C^{[t]}(\bm p) = \min_{n=1,\cdots, N-1}{C^{[t]} (p_n, p_{n+1})}. \label{eqn:capacity}
\end{equation}

Since AANETs have dynamic typologies, a communication link breaks when the transmitter or receiver move out of each other's communication range. To reflect routing stability, we consider the path lifetime as the third metric, which is defined as the duration when every transmitter-receiver pair along the path has direct visibility. A long path lifetime reflects a more stable path, avoiding rerouting. Specifically, the lifetime of path $\bm p$ starting from TS $t$ can be formulated as
\begin{equation}
L^{[t]}(\bm p) = \min_{n=1,\cdots, N-1}{L^{[t]}(p_n, p_{n+1})}. \label{eqn:lifetime}
\end{equation}
where $L^{[t]}(i, j)$ denote the lifetime of link $i\to j$, which is the duration of transmitter $i$ and receiver $j$ being within the communications range, i.e.,
\begin{equation}
L^{[t]}(i,j) = \max_{\Delta t} \left\{ \Delta t \!~ \Big\vert \!~ d^{[t + \Delta t]}(i, j) \leq d_{\sf th} \left(h_i^{[t + \Delta t]}, h_j^{[t + \Delta t]}\right)  \right\}. \label{eqn:linkL}
\end{equation}
We can observe that the link lifetime $L^{[t]}(i,j)$ depends on the future distance between $i$ and $j$ in each time stamp, which further depends on the coordinates of nodes $i$ and $j$ in each future TS, i.e., $\big\{\bm s_{i}^{[t + \Delta t]}\big\}_{t \geq 0}$ and $\big\{\bm s_{j}^{[t + \Delta t]}\big\}_{t \geq 0}$. In practice, the future positions of an airplane can be obtained from its pre-planned trajectory.\footnote{When there is no pre-planned trajectory, the future positions  can be predicted based on the current coordinates, speeds and headings of nodes $i$ and $j$, assuming that they fly at a constant altitude, speed as well as heading.} Then, the distance $d^{[t + \Delta t]}(i, j)$ in each future TS can be calculated for checking whether the maximum communication range is reached, and hence the link lifetime $L^{[t]}(i, j)$ can be calculated effortlessly, by evaluating \eqref{eqn:linkL}  using the classic bi-section search method.

\section{Single-Objective Routing}
In this section, we first formulate a single-objective routing problem aimed at minimizing the E2E delay and then propose the corresponding DL-aided routing algorithm.
\subsection{Problem Formulation}
Let $\bm p^*_{i_{\sf s}}$ denote the optimal path minimizing the E2E delay between an arbitrary source node $i_{\sf s}$ and the destination node $i_{\sf d}$ at an arbitrary TS $t$. The routing optimization problem minimizing the E2E delay can be formulated as
\begin{subequations}
\begin{align}
\mathsf{P}^{[t]}_D:\quad\min_{\bm p} ~& D^{[t]} (\bm p) = \sum_{n=1}^{N - 1} D^{[t]} (p_n, p_{n+1}) \label{eqn:routing}\\
s.t.~& 
d^{[t]}(p_n, p_{n+1}) \leq d_{\sf th}(h_{p_n}^{[t]}, h_{p_{n+1}}^{[t]}), \nonumber  \\
& \phantom{d^{[t]}(p_n, p_{n+1}) \leq}~\forall n=1, \cdots, N-1  \label{eqn:con1}  \\
&  p_n \in {\mathcal{N}^{[t]}}, ~\forall n=1, \cdots, N  \label{eqn:con2}\\
&  p_1 = i_{\sf s}, ~ p_{N} = i_{\sf d}. \label{eqn:con3}
\end{align}
\end{subequations}
which can be regarded as a shortest path problem by treating the link delay as the ``distance". However, conventional graph-based methods, such as Dijkstra’s algorithm, are not applicable in practice due to the following reasons. To solve a shortest path problem using Dijkstra’s algorithm, the global knowledge regarding the graph is required. Specifically, the whole network topology, the propagation and transmission delay between every pair of airplanes, and the queuing delay at each airplane should be known. This requires a centralized controller collecting all the required information, which goes against the self-organized nature of AANETs.  Moreover, due to the high-dynamic network topology of AANETs, the minimum-delay path changes rapidly over time, and hence Dijkstra's algorithm has to be re-run frequently in order to keep the minimum-delay path up-to-date. Consequently, the required global information should be frequently refreshed to the controller, which imposes excessive signaling overhead. Therefore, our goal is to solve problem~$\mathsf P^{[t]}_{D}$  relying solely on local information in a distributed manner.

In the following, we assume that each node is aware of its own position, the positions of its neighbors as well as of the destination, and then invoke DL for finding the optimal path. 

\subsection{DL-Aided Minimum Delay Routing Using Local Information}
\subsubsection{Optimal Substructure}
\begin{figure}[!htb]
	\centering
	\includegraphics[width=0.45\textwidth]{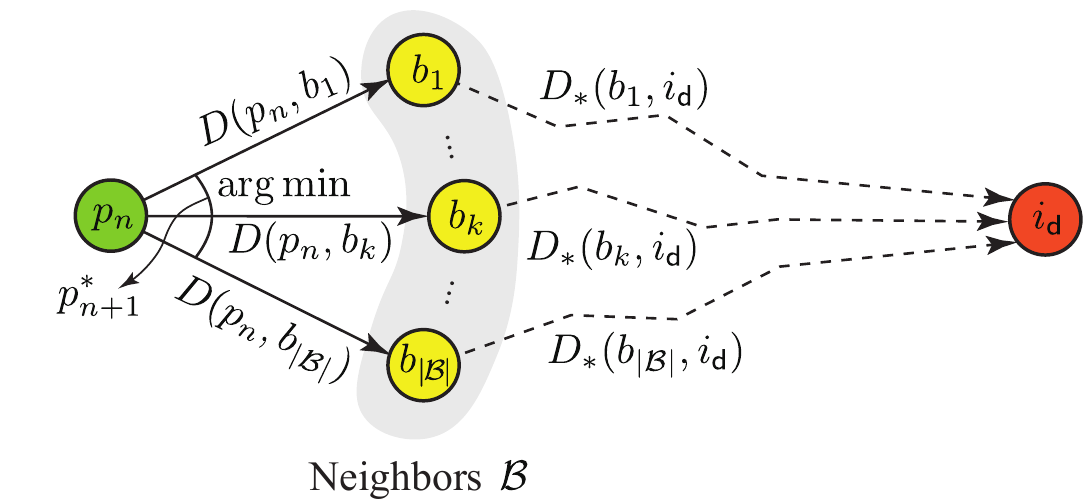}
	\caption{Finding the optimal next hop using the optimal substructure. A straight line indicates a direct communication link and a polyline indicates a minimum-delay path between the two nodes it connects. For notational simplicity, we omit subscript and superscript of ``$\mathcal{B}^{[t]}_{p_n}$", and the parentesis of ``$b_k(p_n)$" }
	\label{fig:opt}
\end{figure}

Problem $\mathsf P^{[t]}_{D}$ has an optimal substructure as shown in Fig.~\ref{fig:opt}. 
Specifically, assume that the packet is currently located at node $p_n$. Let $j\in \mathcal{B}_{p_n}^{[t]}$ denote the index of $p_n$'s neighbor and let $D_*^{[t]}(j, i_{\sf d})$ denote the minimum delay from node $j$ to the destination $i_{\sf d}$. Then, the optimal next hop minimizing the overall delay from the current forwarding node $p_n$ to the destination $i_{\sf d}$ can be characterized by
\begin{equation}
p_{n+1}^* = \arg\min_{j\in \mathcal{B}_{p_n}^{[t]}} \left\{ D^{[t]}(p_n, j) + [1 - I_{\sf d}(j)] D_{*}^{[t]}(j, i_{\sf d}) \right\} \label{eqn:opt},
\end{equation}
where $I_{\sf d} (j) = 1$ if $j=i_{\sf d}$, and $I_{\sf d}(j) = 0$ otherwise. Relying on the optimal substructure, the optimal next hop can be determined when  $D^{[t]}(p_n, j)$ and $D_*^{[t]}(j, i_{\sf d})$ for $j\in \mathcal{B}_{p_n}^{[t]}$ are known. Consequently, the optimal path $\bm p^*_{i_{\sf s}}$ can be obtained by determining the optimal next hop one by one until the packet reaches its destination $i_{\sf d}$.

Since $D^{[t]}(p_n, j)$ can be readily measured by the forwarding node, the optimal next hop can be determined once the minimum delay $D_*^{[t]}(j, i_{\sf d})$ from each neighbor $j\in\mathcal{B}_{p_n}^{[t]}$ to the destination $i_{\sf s}$ is obtained.  Recalling that the goal is to determine the optimal next hop relying solely on local information, it can be accomplished if we can infer $D_*^{[t]}(j, i_{\sf d})$ from local information. In practice, $D_*^{[t]}(j, i_{\sf d})$ depends on the global network topology, which is strongly correlated with the local topology due to the regular flight trajectory and schedule. This motivates us to use a DNN for learning the relationship between the local geographic information observed by the forwarding node and the minimum delay from each neighbor to the destination as follows. 

\subsubsection{Structure of the DNN}
We first specify the input and output of the DNN. To reflect the local topology observed by the forwarding node $i$, the input feature of the DNN should naturally include the position of node $i$ and the positions of its neighbors $\mathcal{B}_i^{[t]}$. However, the number of neighbors is in general different for each node $i \in \mathcal{N}^{[t]}$, while the DNN's input dimension is fixed. If each forwarding node sends its packets to a neighbor that is far away from the destination, then the overall number of hops required for reaching the destination become excessive. Moreover, the computational complexity increases with the number of neighbors. Therefore, it is reasonable to rank the neighbors according to their geographic distance from the destination in ascending order and only consider the top-$K$ ranked neighbors, i.e., $b_{1}(i), \cdots, b_K(i)$. By further including the position of the destination $i_{\sf d}$, the input feature of the DNN can be finally formulated as
\begin{equation}
\bm x_{i}^{[t]} = \left[\bm s_{i}^{[t]}, \bm s_{b_1(i)}^{[t]}, \cdots, \bm s_{b_K(i)}^{[t]}, \bm s_{i_{\sf d}}^{[t]}\right] \in \mathbb{R}^{3(K+2)}, \label{eqn:x1}
\end{equation}
Since the number of neighbors may be lower than $K$ for some nodes, we manually set $\bm s_{b_k(i)}^{[t]} = \bm 0$ if $k > \big\vert\mathcal B_{i}^{[t]}\big\vert$.

Recall that the DNN is used for learning the relationship between the local geographic information and the minimum delay from each neighbor to the destination. Therefore, the desired output of the DNN is designed as
\begin{equation}
\bm y_{i}^{[t]} = \left[D_*^{[t]}(b_1(i), i_{\sf d}), \cdots, D_*^{[t]}(b_K(i), i_{\sf d})\right] \in \mathbb{R}^{K},
\end{equation}
where again, we only consider the top-$K$ ranked neighbors for fixing the output dimension of the DNN and manually set $D_*^{[t]}(b_k(i), i_{\sf d}) = \sf inf$\footnote{$\sf inf$ is defined as a very large number, e.g., $\mathsf{inf} = 10^8$.} for $k > \big|\mathcal B_{i}^{[t]}\big|$.  

\begin{figure}[!htb]
	\centering
	\includegraphics[width=0.5\textwidth]{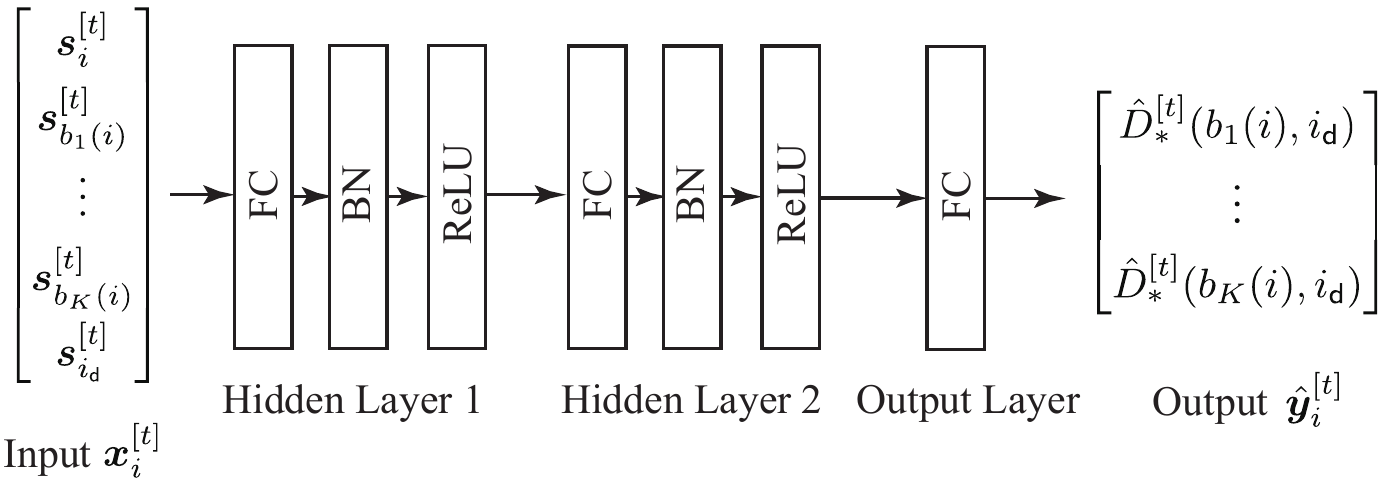}
	\caption{DNN architecture for minimum delay routing, where ``FC" represents the fully-connected layer, ``BN" stands for batch normalization, and ``ReLU" is the rectified linear unit.}
	\label{fig:dnn}
\end{figure}

The structure of the DNN is shown in Fig.~\ref{fig:dnn}. Since each dimension of the input has different units, e.g. [$^\circ$] for the longitude and latitude while [km] for the altitude, we use batch normalization~\cite{batch} for normalizing each dimension across the samples in a mini-batch to have zero mean and unit variance. 

Let $\bm \theta$ denote the collection of unknown parameters to be trained. The DNN can be expressed as a function of $\bm x_i^{[t]}$ with parameter $\bm \theta$ as $\bm f(\bm x_{i}^{[t]}; \bm \theta)$, whose output is the estimate of $\bm y_i^{[t]}$ formulated as
\begin{equation}
\hat{\bm y}_{i}^{[t]} = \bm f(\bm x_{i}^{[t]}; \bm \theta) = \left[\hat D^{[t]}_*(b_1(i), i_{\sf d}), \cdots, \hat D_*^{[t]}(b_K(i), i_{\sf d})\right],
\end{equation}
where $\hat D_*^{[t]}(b_k(i), i)$ denotes the estimate of $D_*^{[t]}(b_k(i), i)$, and again, we manually set $\hat D_*^{[t]}(b_k(i), i_{\sf d}) = \sf inf$ for $k > \big|\mathcal B_{i}^{[t]}\big|$.

The expression of $\bm f (\cdot; \bm \theta)$ is determined by cascading the expression of each layer in the DNN as 
\begin{equation}
\hat {\bm y} = \bm f({\bm x}; \bm \theta) = \bm f_K \left( \cdots \left( \bm f_k \left( \cdots \bm f_1 \left({\bm x}; \bm\theta_1\right) \cdots\right); \bm \theta_k \right) \cdots; \bm \theta_K\right),
\end{equation}
where $\bm f_k( \cdot; \bm \theta_k)$ represents the $k$th layer of the DNN with parameter $\bm \theta_k$.

Specifically, the relationship between the output ${\bm y}_{\sf out}$ and input ${\bm x}_{\sf in}$ of the  fully-connected layer is expressed as
\begin{equation}
{\bm y}_{\sf out} = \bm f_{\sf FC}({\bm x}_{\sf in}; \bm W, \bm b) = \bm W {\bm x}_{\sf in} + \bm b,
\end{equation}
where $\bm W$ and $\bm b$ denote the weight and bias parameters, respectively. The batch normalization layer can be represented by
\begin{equation}
{\bm y}_{\sf out} = \bm f_{\sf BN}({\bm x}_{\sf in};\bm \gamma,\bm \beta) = \bm \gamma \circ \frac{{\bm x}_{\sf in} - \bm \mu}{\sqrt{\bm \sigma^2 + \epsilon}} + \bm \beta,
\end{equation}
where ``$\circ$" denotes the element-wise multiplication, $\bm \gamma$ and $\bm  \beta$ are the scale and shift parameters, respectively, and $\epsilon$ is a small constant introduced for ensuring numerical stability. During training, $\bm \mu$ and $\bm  \sigma^2$ represent the empirical mean and variance of the input mini-batch, respectively. During testing (i.e., inference), $\bm \mu$ and $\bm \sigma^2$ represent the population mean and variance of the entire training set~\cite{batch}. The rectified linear unit (ReLU) layer implements the following operation
\begin{equation}
{\bm y}_{\sf out} = \bm f_{\sf ReLU} ({\bm x}_{\sf in}) = \max \{{\bm x}_{\sf in}, \bm 0 \},
\end{equation}
where the $\max$ operation is carried out in an element-wise manner.

\subsubsection{Offline Training}
Since the takeoff times as well as the trajectories of commercial passenger airplanes are pre-planned and normally repeated on a weekly basis, the flight data on the same day of different weeks are similar. Therefore, we can train the DNN based on historical flight data or on the pre-planned flight trajectories in an offline manner.

The goal of training is to  minimize the loss function composed of the mean squared error between the actual output $\hat{\bm y}_i^{[t]} = \bm f(\bm x_{i}^{[t]}; \bm \theta)$ and the desired output  $\bm y_i^{[t]}$ of the DNN, yielding:
\begin{equation}
{\sf P_{loss}:}\quad \min_{\bm \theta}~ \frac{1}{N_{\sf total}} \sum_{t\in \mathcal{T}}\sum_{i \in \mathcal{N}^{[t]}} \left\Vert \bm y_{i}^{[t]} - \bm f(\bm x_{i}^{[t]}; \bm \theta)  \right\Vert^2,
\end{equation}
where $\mathcal{T}$ denotes the set of TSs of the historical flight data and $N_{\sf total}$ is the number of training samples.

In the following, we first generate the training samples, including the input $\bm x_i^{[t]}$ and the desired output $\bm y_i^{[t]}$ for $t\in \mathcal{T}$ and $i \in \mathcal{N}^{[t]}$. For each TS $t$ in the historical dataset, we can retrieve the position of each node from the flight trajectories and create a snapshot of the whole network topology. Consequently, for each node $i\in \mathcal{N}^{[t]}$, we can obtain the DNN's input $\bm x_i^{[t]}$ from the position of each node. 

To obtain the desired output $\bm y_i^{[t]}$ as the training label, we have to compute the minimum delay from each neighbor of node $i$ to the destination node, i.e., $\hat D_*^{[t]}(j, i_{\sf d})$ for $\forall j\in \mathcal{B}_{i}^{[t]}$. Since the position of each node can be retrieved from the historical flight data, the transmission delay and propagation delay of each link can be readily calculated based on the expressions given after~\eqref{eqn:Dlink}. As for the queuing delay, since we aim to train the DNN for exploiting the correlation between local and global network topology, which is independent from the traffic load, we assume that the queuing delay is identical and constant among all the nodes during training. In this way, the total queuing delay is determined by the number of hops in the path so that the network topology information is implicitly reflected. Then, the delay of every link in the network can be calculated by~\eqref{eqn:Dlink}. Consequently, we can obtain the minimum delay $D_*^{[t]}(j, i_{\sf d})$ for $\forall j\in \mathcal{B}_{i}^{[t]}$ by setting $i_{\sf s} = j$ and solving problem $\mathsf P^{[t]}_D$ using any shortest path algorithm.

Upon obtaining the training set, we can train the DNN by solving problem $\mathsf P^{[t]}_{\sf loss}$ using stochastic gradient descent algorithms, such as the Adam method~\cite{adam}. 
We note that the DNN is not designed and trained for a specific node. Instead, it can be used by all the forwarding nodes during the time window $\mathcal{T}$ in online testing. In fact, different forwarding nodes are distinguished by their local geographic information $\bm x_i^{[t]}$ and the desired DNN's input-output pairs of all nodes $\mathcal{N}_t$ during $\mathcal{T}$, i.e., $(\bm x_i^{[t]}, \bm y_i^{[t]})$ for all $i\in \mathcal{N}_t$ and $t\in \mathcal{T}$, are included in the training set. Therefore, once the DNN becomes sufficiently well-trained, all the forwarding nodes are able to obtain the required information for determining their own optimal next hop from the same DNN by inputting their own local geographic observation. 
	
The benefit of all nodes using the same DNN is that it enables a certain degree of parameter sharing among different nodes, which improves the learning efficiency and scalability. For example, when two nodes are close, they share a similar set of neighbors and hence provide similar input for the DNN. Moreover, the outputs of their DNNs (i.e., the minimum delay from the neighbors to the destination) should  also be similar, which suggests that the DNNs of these pair of nodes should be similar. This can be naturally achieved if the nodes share the same DNN.

\subsubsection{On-line Decision Without Feedback}
Once the DNN has become sufficiently well-trained, the DNN is copied to each airplane in support of online routing decisions. 

Assume that the packet is currently located at node $p_n$. We first specify the next-hop candidates, namely the set of nodes where the next hop of $p_n$ is chosen from. Since the input and output dimensions of the DNN are fixed, the next-hop candidates are restricted the top-$K$ ranked neighbors of $p_n$.\footnote{Although only considering the top-$K$ ranked neighbors may potentially degrade the performance, when $K$ is sufficiently large, the performance degradation is negligible as shown in our simulation results in Section~V.}

Moreover, to avoid loops in the path, the nodes that  already exist in the path should be excluded. Therefore, the next-hop candidate set of $p_n$ is specified as
\begin{equation}
\mathcal{C}_{p_n}^{[t]} = \{b_{1}(p_n), \cdots, b_K(p_n)\}\backslash \{p_1, \cdots, p_n\}
\end{equation}

By observing the local geographic information, node $p_n$ can calculate the delay from itself to each of its next-hop candidates, i.e., $D^{[t]}(p_n, j)$ for $j\in\mathcal{C}_{p_n}^{[t]}$, and gathers its input features $\bm x_{p_n}^{[t]}$ for the DNN. Then, from the output of the DNN, the forwarding node $p_n$ can obtain the estimate of the minimum delay from each of its next-hop candidates to the destination, i.e., $\hat D_*^{[t]}(j, i_{\sf d})$ for $j\in\mathcal{C}_{p_n}^{[t]}$. 

According to the optimal substructure \eqref{eqn:opt}, a straightforward way of determining the next hop using the DNN is by evaluating
\begin{equation}
p_{n+1} = \arg\min_{j \in \mathcal{C}_{p_n}^{[t]} }  \left\{ D^{[t]} (p_n, j) + \left[1 - I_{\sf d}(j)\right]  \hat D_*^{[t]}(j, i_{\sf d})  \right\}. \label{eqn:nof}
\end{equation}

Recall that the training of DNN treats the queuing delay as an identical
constant for each node, while in reality the queuing delay varies among nodes due to different traffic load of each node. Moreover, although the weekly flight trajectories and schedules are similar, there exist a certain mismatch between the historical flight data and the current flight data. Hence, there may exist errors between the minimum delay $D_*^{[t]}(j, i_{\sf d})$ and its estimate $\hat D_*^{[t]}(j, i_{\sf d})$, which degrades the online routing performance. In the following, we introduce a feedback mechanism for improving the online routing decisions.

\subsubsection{On-line Decision with Feedback} 
Let $j_1 \in \mathcal{C}_{p_n}$ denote the index of the next-hop candidate of $p_n$ and $j_2 \in \mathcal{C}_{j_1}$ denote the index of the next-hop candidates of $j_1$, i.e., the next-but-one or \emph{next-2-hop} candidate of $p_n$. Instead of directly computing $\hat D_*^{[t]}(j_1, i_{\sf d})$ by inputting the geographic information $\bm x_{p_n}^{[t]}$ observed by the forwarding node $p_n$ into the DNN, we let each next-hop candidate $j_1 \in \mathcal{C}_{p_n}$  compute $\hat D_*^{[t]}(j_2, i_{\sf d})$ by inputting their local geographic observation $\bm x_{j_1}^{[t]}$ into the DNN and measure the link delay $D^{[t]}(j_1, j_2)$. Then, by letting $j_1 \in \mathcal{C}_{p_n}$ report $D^{[t]}(j_1, j_2)$ and $\hat D_*^{[t]}(j_2, i_{\sf d})$ back to the forwarding node $p_n$, the minimum delay from $j_1$ to the destination can be estimated based on the optimal substructure \eqref{eqn:opt} using $D^{[t]}(j_1, j_2)$ and $\hat D_*^{[t]}(j_2, i_{\sf d})$.  In the sequel, we provide an example snapshot for illustrating the on-line routing decision process relying on the feedback mechanism.
\begin{figure}
	\centering	
	\subfigure[Without feedback]{
		\label{fig:opt1} 
		\includegraphics[height=0.22\textwidth]{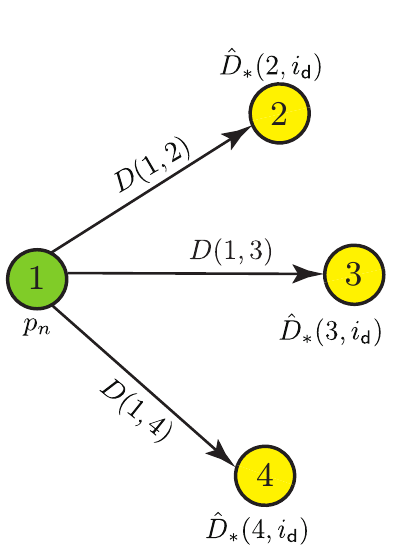}}
	\subfigure[With feedback]{
		\label{fig:opt2} 
		\includegraphics[height=0.22\textwidth]{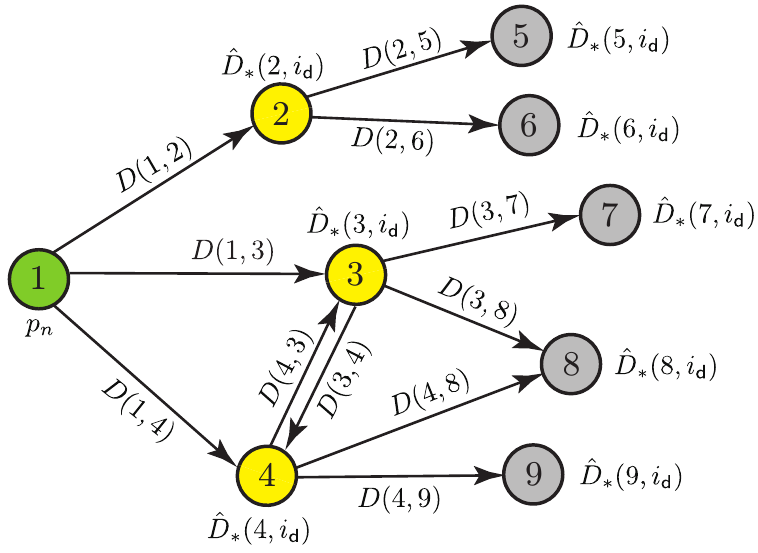}}
	\caption{Comparison of the available information at the forwarding node for determining the next hop.}
	\label{fig:compare}
\end{figure}

In Fig.~\ref{fig:compare}, we first compare the information for determining the next hop acquired both with and without the feedback mechanism in an example snapshot. Let us assume that the packet is currently located at node~$1$, i.e., $p_n = 1$, and the maximum number of neighbors to be considered as candidates is $K=3$. As shown in Fig.~\ref{fig:opt1}, the next-hop candidates of node $1$ are node $2$, $3$ and $4$. The next-hop candidates of nodes $2$, $3$ and $4$ are $\{5,6\}$, $\{4, 7, 8\}$ and $\{3, 8,9\}$, respectively, as shown in Fig.~\ref{fig:opt2}. 

Without the feedback mechanism, node $1$ first observes its local geographic information $\bm x_1^{[t]}$, which is then used as the input of the DNN for estimating the minimum delays from nodes $\{2,3,4\}$ to the destination, i.e., 
\begin{equation}
[\hat{D}_*^{[t]}(2, i_{\sf d}), \hat{D}_*^{[t]}(3, i_{\sf d}), \hat{D}_*^{[t]}(4, i_{\sf d})] = \bm f(\bm x_{1}^{[t]}; \bm \theta). \label{eqn:DNNest}
\end{equation}
Meanwhile, node $1$ measures the link delays from itself to nodes $\{2, 3, 4\}$, i.e., $[D^{[t]}(1, 2), D^{[t]}(1, 3),$ $ D^{[t]}(1, 4)]$. According to~\eqref{eqn:nof}, the next hop can be determined by
\begin{equation}
p^{n+1} = \arg\min_{j\in\{2,3,4\}} \left\{D^{[t]}(1, j) + \left[1 - I_{\sf d}(j)\right] \hat D_*^{[t]}(j, i_{\rm d})\right\}. \label{eqn:pn}
\end{equation}

After introducing the feedback mechanism, the minimum delay from the next-hop candidates $\{2,3,4\}$ can be estimated in the following manner instead. Specifically, nodes $\{2,3,4\}$ first observe their local geographic information $\bm x_2^{[t]}$, $\bm x_3^{[t]}$ and $\bm x_4^{[t]}$, respectively. The minimum delays from all the next-2-hop candidates to the destination can be estimated  based on the DNN by inputting $\bm x_2^{[t]}$, $\bm x_3^{[t]}$ and $\bm x_4^{[t]}$, respectively. For example, the minimum delays from nodes $\{5, 6\}$ to the destination can be estimated by
\begin{equation}
\left[\hat{D}_*^{[t]}(5, i_{\sf d}), \hat{D}_*^{[t]}(6, i_{\sf d})\right] = \bm f(\bm x_{2}^{[t]}; \bm \theta)[:2],
\end{equation}
where $\bm y[:n]$ denotes the first $n$ elements of vector $\bm y$. Similarly, we can obtain
\begin{subequations}
\begin{align}
\left[\hat{D}_*^{[t]}(4, i_{\sf d}), \hat{D}_*^{[t]}(7, i_{\sf d}), \hat{D}_*^{[t]}(8, i_{\sf d})\right] & = \bm f(\bm x_{3}^{[t]}; \bm \theta)\\
\left[\hat{D}_*^{[t]}(3, i_{\sf d}), \hat{D}_*^{[t]}(8, i_{\sf d}), \hat{D}_*^{[t]}(9, i_{\sf d})\right] & = \bm f(\bm x_{4}^{[t]}; \bm \theta) .
\end{align}
\end{subequations}
Meanwhile, nodes $2$, $3$ and $4$ measure the link delays $\{D^{[t]}(2,5), D^{[t]}(2,6)\}$, $\{D^{[t]}(3,4), D^{[t]}(3,7),$ $D^{[t]}(3,8)\}$ and $\{D^{[t]}(4,3), D^{[t]}(4,8), D^{[t]}(4,9)\}$, respectively.  Then, with the aid of the feedback mechanism, node $1$ now becomes aware of the link delays between itself to its next-hop candidates as well as the link delays between the next-hop candidates and the next-2-hop candidates, as shown Fig.~\ref{fig:opt2}. This allows the forwarding node to select the next-hop more wisely, since more link information (i.e., the queuing delay of next-hop candidates as well as the propagation and transmission delay from each next-hop candidate to the corresponding next-2-hop candidates) can now be taken into consideration by recursively exploiting the optimal substructure. 
\begin{figure}
	\centering	
	\subfigure[Compute $\tilde D_*^{[t]}(2, i_{\sf d})$]{
		\label{fig:a} 
		\includegraphics[height=0.15\textwidth]{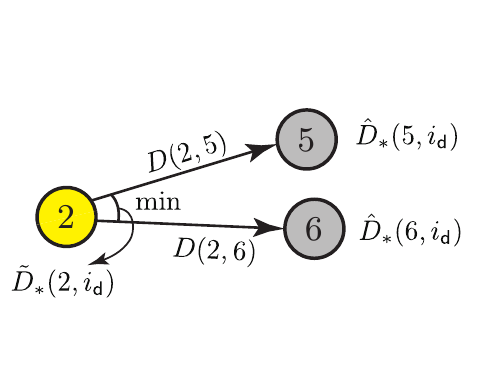}}
	\subfigure[Compute $\tilde D_*^{[t]}(3, i_{\sf d})$]{
		\label{fig:b} 
		\includegraphics[height=0.15\textwidth]{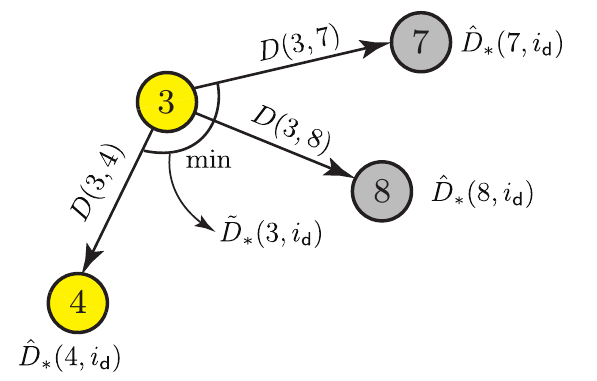}}
	\subfigure[Compute $\tilde D_*^{[t]}(4, i_{\sf d})$]{
		\label{fig:c} 
		\includegraphics[height=0.15\textwidth]{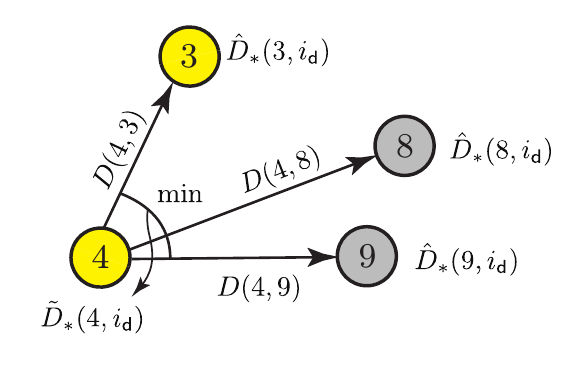}}
	\caption{Compute $\tilde D_*^{[t]}(j_1, i_{\sf d})$, $j_1\in  \mathcal{C}_{p_n}^{[t]}$ for the first round to determine the update order for decoupling the mutual dependence.}
	\label{fig:t}
	\vspace{-5mm}
\end{figure}

Consequently, the minimum delay from nodes $\{2, 3, 4\}$ to the destination can be estimated using the optimal substructure instead of using~\eqref{eqn:DNNest}, as illustrated in Fig.~\ref{fig:t}. This is formulated as follows
\begin{align}
\tilde D_*^{[t]} (2, i_{\sf d}) & = \min_{j_2\in\{5, 6\}} D^{[t]}(2, j_2) + \hat D_*^{[t]}(j_2, i_{\rm d}) \\
\tilde D_*^{[t]} (3, i_{\sf d}) & = \min_{j_2\in\{4, 7, 8\}} D^{[t]}(3, j_2) + \hat D_*^{[t]}(j_2, i_{\rm d}) \label{eqn:3}\\
\tilde D_*^{[t]} (4, i_{\sf d}) & = \min_{j_2\in\{3, 8, 9\}} D^{[t]}(4, j_2) + \hat D_*^{[t]}(j_2, i_{\rm d}). \label{eqn:4}
\end{align}

It is worth noting that node $3$ and node $4$ are not only the next-hop candidates of the forwarding node $1$, but they are also  the next-hop candidates of each other. From \eqref{eqn:3} and \eqref{eqn:4}, we can see that the computation of $\tilde D_*^{[t]} (3, i_{\sf d})$ and $\tilde D_*^{[t]} (4, i_{\sf d})$ depends on the estimate $\hat D_*^{[t]} (4, i_{\sf d})$ and $\hat D_*^{[t]} (3, i_{\sf d})$, respectively. This suggests that the estimate $\tilde D_*^{[t]} (4, i_{\sf d})$ can be computed by 
\begin{align}
\tilde D_*^{[t]} (4, i_{\sf d}) = \min \Big\{&D^{[t]}(4, 3) + \tilde D_*^{[t]}(3, i_{\rm d}), \nonumber\\
&  D^{[t]}(4, 8) + \hat D_*^{[t]}(8, i_{\rm d}), \nonumber \\
& D^{[t]}(4, 9) + \hat D_*^{[t]}(9, i_{\rm d})\Big\} \label{eqn:42}
\end{align}
instead, where $\tilde D_*^{[t]}(3, i_{\rm d})$ is computed by~\eqref{eqn:3}. Compared with using \eqref{eqn:4}, the real-time link delays of node $3$ to its next-hop candidates, i.e., $\{D(3, 4), D(3, 8), D(3, 9)\}$,  are also taken into account when computing $\tilde D_*^{[t]}(4, i_{\rm d})$ using~\eqref{eqn:42} and~\eqref{eqn:3}. Alternatively, $\tilde D_*^{[t]} (3, i_{\sf d})$ can be also computed by
\begin{align}
\tilde D_*^{[t]} (3, i_{\sf d}) = \min \Big\{ & D^{[t]}(3, 4) + \tilde D_*^{[t]}(4, i_{\rm d}), \nonumber\\ 
& D^{[t]}(3, 8) + \hat D_*^{[t]}(8, i_{\rm d}), \nonumber\\
& D^{[t]}(3, 9) + \hat D_*^{[t]}(9, i_{\rm d})\Big\} \label{eqn:32}
\end{align}
in a similar manner to~\eqref{eqn:42}. Consequently, the computation of $\tilde D_*^{[t]} (3, i_{\sf d})$ and $\tilde D_*^{[t]} (4, i_{\sf d})$ are mutually dependent on each other, as shown in~\eqref{eqn:42} and~\eqref{eqn:32}. 

The mutual dependence arises from the fact that the packet can be sent between nodes $3$ and $4$ in either direction. However, link $4\to 3$ and link $3\to 4$ cannot co-exist in the optimal path. In the following, we decouple the mutual dependence by specifying the direction of the link between the mutually dependent candidates and by specifying the order of mutually dependent candidates for the minimum delay estimation.  

In general, we define 
\begin{equation}
\mathcal{M}^{[t]}_{p_n} \triangleq \left\{j_2 \left\vert \left(j_2\in \mathcal{C}_{p_n}^{[t]}\right) \land \left(\exists j_1   \in \mathcal{C}_{p_n}^{[t]}~s.t.~j_2\in\mathcal{C}_{j_1}^{[t]}\right) \right.\!\right\} \!\!\! \label{eqn:M}
\end{equation} 
as the \emph{mutual candidate set}, which contains the specific next-hop candidates of $p_n$ that are also the next-2-hop candidates of $p_n$. Since the delay to the destination decreases along the optimal path, the packet should only be sent to a node having a lower delay to the destination than that of the forwarding node. Therefore, the nodes in $\mathcal{M}_{p_n}^{[t]}$ are sorted according to the estimated minimum delay to the destination in ascending order. Specifically, let $m_k$ denote the $k$th element of $\mathcal{M}_{p_n}^{[t]}$. We have $\tilde D_*^{[t]}(m_1, i_{\sf d}) \leq \tilde D_*^{[t]}(m_2, i_{\sf d}) \leq \cdots \leq \tilde D_*^{[t]}(m_{|\mathcal{M}_{p_n}^{[t]}|}, i_{\sf d})$.

\begin{figure}
	\centering
	\includegraphics[height=0.24\textwidth]{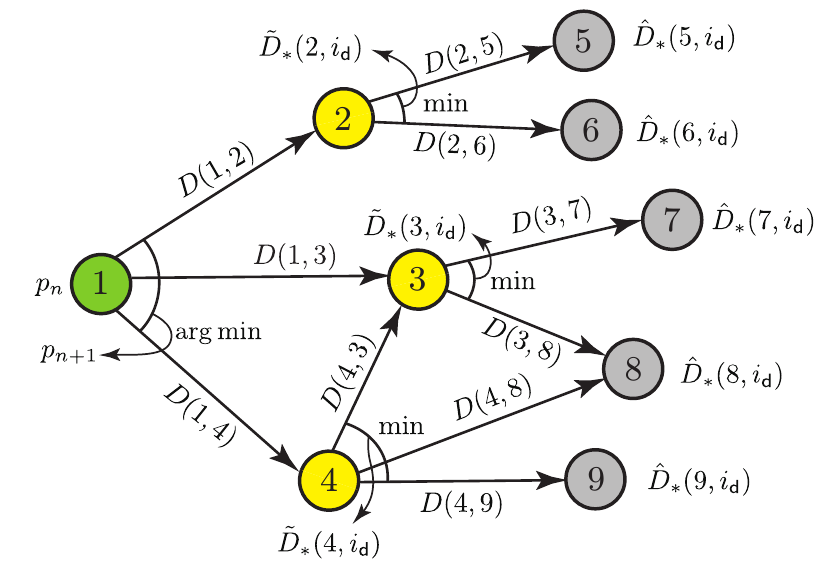}
	\caption{The computation graph for determining the next hop $p_{n+1}$. }
	\label{fig:second}
	\vspace{-2mm}
\end{figure}

For the example considered, assume that we have $\tilde D_*^{[t]} (3, i_{\sf d}) < \tilde D_*^{[t]} (4, i_{\sf d})$ based on the value computed from~\eqref{eqn:3} and~\eqref{eqn:4}. Then, the link $3 \to 4$ is less likely to exist in the optimal path. Therefore, we recompute the value of $\tilde D_*^{[t]} (3, i_{\sf d})$ and $\tilde D_*^{[t]} (4, i_{\sf d})$ as follows
\begin{align}
\tilde D_*^{[t]} (3, i_{\sf d})& = \min_{j_2\in\{7, 8\}} \left\{ D^{[t]}(3, j_2) + \hat D_*^{[t]}(j_2, i_{\rm d})\right\} \label{eqn:33} \\
\tilde D_*^{[t]} (4, i_{\sf d}) & = \min \Big\{D^{[t]}(4, 3) + \tilde D_*^{[t]}(3, i_{\rm d}), \nonumber\\
& \hphantom{=\min \Big\{~} D^{[t]}(4, 8) + \hat D_*^{[t]}(8, i_{\rm d}), \nonumber\\
& \hphantom{=\min \Big\{~} D^{[t]}(4, 9) + \hat D_*^{[t]}(9, i_{\rm d}) \Big\}, \label{eqn:43}
\end{align}
where node $4$ is no longer considered as the next-hop candidate of node $3$ and the computation of $\tilde D_*^{[t]}(4, i_{\rm d})$  depends on the value of $\tilde D_*^{[t]}(3, i_{\rm d})$.  Finally the next hop can be determined by
\begin{equation}
p_{n+1} = \arg\min_{j_1 \in \{2,3,4\}}  \left\{D^{[t]}(1, j_1) + \left[1 - I_{\sf d}(j)\right] \tilde D_*^{[t]}(j_1, i_{\rm d})\right\}.
\end{equation}
The computation graph constructed for determining the next hop $p_{n+1}$ is summarized in Fig.~\ref{fig:second} after decoupling the mutual dependence.

In general, the computation of $\tilde D_*^{[t]}(j_1, i_{\sf d})$ for $j_1\in \mathcal{C}_{p_n}$ is formulated as follows. We first initialize the value of $\tilde D_*^{[t]}(j_2, i_{\sf d})$ for each  $j_2 \in \mathcal{C}_{j_1}$ and each $j_1 \in \mathcal{C}_{p_n}$ by $\tilde D_*^{[t]}(j_2, i_{\sf d}) \leftarrow \hat D_*^{[t]}(j_2, i_{\sf d})$.

Next, we compute the value of $\tilde D_*^{[t]}(j_1, i_{\sf d})$  for each $j_1 \in \mathcal{C}_{p_n}$ in the first round as
\begin{align}
\tilde D_*^{[t]}(j_1, i_{\sf d}) \leftarrow & \left[1 - I_{\sf d}(j_1)\right]\cdot \min_{j_2 \in \mathcal{C}_{j_1}^{[t]} } \Big\{ D^{[t]} (j_1, j_2)\nonumber\\
& +  \left[1 - I_{\sf d}(j_2)\right]  \hat D_*^{[t]}(j_2, i_{\sf d}) \Big\}.  \label{eqn:1st}
\end{align}

After that, we find the mutual candidate set $\mathcal{M}_{p_n}^{[t]}$ defined in~\eqref{eqn:M} and sort its elements according to $\tilde D^{[t]}_*(m_1, i_{\sf d}) \leq \tilde D^{[t]}_*(m_2, i_{\sf d}) \leq \cdots \leq \tilde D^{[t]}_*(m_{|\mathcal{M}_{p_n}^{[t]}|}, i_{\sf d})$ using the value computed by~\eqref{eqn:1st}. Moreover, we define 
\begin{align}
\bar {\mathcal{C}}_{m_k}^{[t]} =  \mathcal{C}_{m_k}^{[t]} - \Big\{ j \Big\vert & \left(j \in \mathcal{C}_{m_k}^{[t]}\right) \land \left(j\in\mathcal{M}_{p_n}^{[t]} \right) \nonumber \\
&  \land \left(\tilde D_*^{[t]}(j, i_{\sf d}) > \tilde D_*^{[t]}(m_k, i_{\sf d})\right) \Big\},  \label{eqn:barC}
\end{align}
which excludes the specific next-hop candidates of $m_k$ that are also in the sorted mutual candidate sets $\mathcal{M}_{p_n}^{[t]}$ and have a higher minimum delay to the destination than that of $m_k$.

Then, we recompute the value of $\tilde D_*^{[t]}(j_1, i_{\sf d})$ for $j_1\in\mathcal{M}_{p_n}^{[t]}$ one by one according the rank of $j_1$ in $\mathcal{M}_{p_n}^{[t]}$. To be more specific, for $k = 1, \cdots, \big\vert\mathcal{M}_{p_n}^{[t]}\big\vert$, we compute
\begin{align}
\tilde D_*^{[t]}(m_k, i_{\sf d}) \leftarrow & \left[1 - I_{\sf d}(m_k)\right]\cdot\min_{j_2 \in \bar{\mathcal{C}}_{m_k}^{[t]}   }   \Big\{ D^{[t]} (m_k, j_2) \nonumber\\
& + \left[1 - I_{\sf d}(j_2)\right]  \tilde D_*^{[t]}(j_2, i_{\sf d})  \Big\}. \label{eqn:2nd}
\end{align}

Finally, the next hop is determined by the optimal substructure as follows
\begin{equation}
\! p_{n+1} = \arg\!\min_{j \in \mathcal{C}_{p_n}^{[t]}}  \left\{ D^{[t]} (p_n, j) + \left[1 - I_{\sf d}(j)\right]  \tilde D_*^{[t]}(j, i_{\sf d})  \right\}. \!\!\!\!
\end{equation}

In Algorithm~\ref{alg:single_DL}, we summarized procedure of the DL-aided routing during the online decision phase.
\begin{algorithm}[!htb]
	\caption{DL-aided routing for minimizing the E2E delay}\
	\label{alg:single_DL}
	\begin{algorithmic}[1]
		\Algtop{$\tt \%~Obtain~the~information~for~determining~the$ $\tt next~hop$}
		\State The forwarding node $p_n$ discovers its next-hop candidates $\mathcal{C}_{p_n}$ and measures the delay from itself to each next-hop candidate $D^{[t]}(p_n, j_1)$, $j_1\in \mathcal{C}_{p_n}$.
		\State Each next-hop candidate $j_1 \in \mathcal{C}_{p_n}$ computes $\hat{D}_*^{[t]}(j_2, i_{\sf d})$ using the DNN, measures $D^{[t]}(j_1, j_2)$ for $j_2\in\mathcal{C}_{j_1}$, and feeds all the information back to node $p_n$. 
		\Algphase{$\tt\%~Computations~at~the~forwarding~node~for$ $\tt~ determining~the~next~hop$}
		\State Initialize $\tilde D_*^{[t]}(j_2, i_{\sf d}) \leftarrow \hat D_*^{[t]}(j_2, i_{\sf d})$ for each $j_2 \in \mathcal{C}_{j_1}$, $j_1 \in \mathcal{C}_{p_n}$.
		\State Update $\tilde D_*^{[t]}(j_1, i_{\sf d})$ using \eqref{eqn:1st} for each $j_1 \in \mathcal{C}^{[t]}_{p_n}$.

		\State Find $\mathcal{M}_{p_n}^{[t]}$ defined by~\eqref{eqn:M} and sort the element of $\mathcal{M}_{p_n}^{[t]}$ by $\tilde D_*^{[t]}(m_1, i_{\sf d}) \leq \tilde D_*^{[t]}(m_2, i_{\sf d}) \leq \cdots \leq \tilde D_*^{[t]}(m_{|\mathcal{M}_{p_n}^{[t]}|}, i_{\sf d})$.
		\For{$k = 1, \cdots, \big\vert\mathcal{M}_{p_n}^{[t]}\big\vert$}
		\State Find $\bar {\mathcal{C}}_{m_k}^{[t]}$ defined by~\eqref{eqn:barC}.
		\State Update $\tilde D_*^{[t]}(m_k, i_{\sf d})$ using~\eqref{eqn:2nd}.
		\EndFor
		\State Transmit the packet to $$p_{n+1} = \arg\min\limits_{j \in \mathcal{C}_{p_n}^{[t]}}  \left\{ D^{[t]} (p_n, j) + \left[1 - I_{\sf d}(j)\right]  \tilde D_*^{[t]}(j, i_{\sf d})  \right\}$$
	\State Set $n\leftarrow n+1$ and repeat steps 1 $\sim$ 9 until the packet reaches the destination.
	\end{algorithmic}
\end{algorithm}

\section{Multi-Objective Routing}
In this section, we extend our DL-aided routing to a challenging multi-objective scenario that considers both the E2E delay as well as the path capacity and path lifetime. We first introduce some basic concepts of MOO and formulate the multi-objective routing problem for simultaneously minimizing the delay, maximizing the capacity and maximizing the path lifetime. Then, we solve the problem using global information for obtaining the true Pareto front, and finally proposes an efficient DL-aided multi-objective routing algorithm that  relies solely on local information for making online routing decisions in practical AANETs.
\subsection{Problem Formulation}
A standard MOO problem can be formulated as 
\begin{align}
\min_{\bm x}~& \bm g(\bm x) = [g_1(\bm x), g_2(\bm x), \cdots, g_M(\bm x)] \\
s.t.~& \bm x \in \mathcal{X},
\end{align}
where we have the following definitions. 
\begin{definition} \rm
	\emph{Pareto dominance}: Given a distinct pair of solutions $\bm x_1, \bm x_2 \in \mathcal{X}$, $\bm x_1$ is said to \emph{dominate} $\bm x_2$, if and only if:
	\begin{enumerate}
		\item $g_m(\bm x_1) \leq g_m(\bm x_2)$ for any $m\in\{1, 2, \cdots, M\}$;
		\item There exist an $m\in\{1, 2, \cdots, M\}$ such that $g_m(\bm x_1) < g_m(\bm x_2)$.
	\end{enumerate}
	Accordingly, in the objective space, $\bm g(\bm x_1)$ is said to \emph{dominate} $\bm g(\bm x_2)$.
\end{definition}

\begin{definition} \rm
	\emph{Pareto optimality}: A solution $\bm x \in \mathcal{X}$ is said to be \emph{Pareto optimal} if and only if there does not exist another point $\bm x' \in \mathcal{X}$ that dominates $\bm x$. Furthermore, all the Pareto-optimal solutions mapped in the objective space constitutes the \emph{Pareto front}.
\end{definition}

\begin{definition} \rm
	\emph{Weak Pareto optimality}: A solution $\bm x \in \mathcal{X}$ is said to be \emph{weak Pareto optimal} if and only if there does not exist another point $\bm x' \in \mathcal{X}$ such that $g_m(\bm x') < g_m(\bm x)$ for any $m\in\{1, 2, \cdots, M\}$.
\end{definition}

\begin{figure}[!htb]
	\centering
	\includegraphics[width=0.3\textwidth]{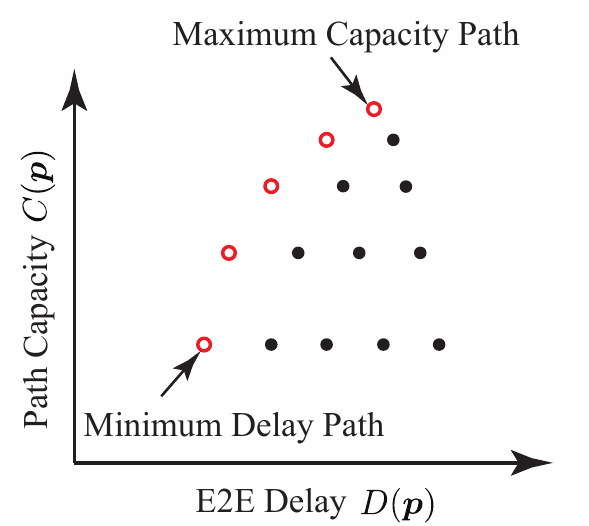}
	\caption{Illustration of multi-objective routing.}
	\label{fig:moo}
\end{figure}
Let us use a toy example to augment the above concepts more concretely. In Fig.~\ref{fig:moo}, we consider a twin-objective routing problem, where the objectives are to simultaneously minimize the E2E delay and maximize the path capacity, i.e., $\min\limits_{\bm p}~ [D(\bm p), -C(\bm p)]$. Each point in Fig.~\ref{fig:moo} represents a possible path $\bm p$ from the source node to the destination node, while the $x$-coordinate and $y$-coordinate of a point indicate the delay and capacity of the path, respectively. Specifically, the black filled circles represent the non-Pareto-optimal solutions (paths), whose capacity can be increased or E2E delay can be reduced without sacrificing the other. By contrast, the red hollow circles represent the Pareto-optimal solutions (paths), which constitute the Pareto front. We can observe that the minimum delay and the maximum capacity cannot be achieved by a single path. Instead, there is a optimal trade-off relationship (characterized by the Pareto front) between the path capacity and E2E delay, i.e., which specifics the minimum increment of E2E delay required for achieving a certain path capacity improvement and vice versa.

Finally, the multi-objective routing problem minimizing the delay, maximizing the capacity and maximizing the path lifetime can be formulated in the following standard form
\begin{subequations}
\begin{align}
\mathsf{P}^{[t]}_{D, C, L}: \quad \min_{\bm p} ~& [D^{[t]} (\bm p), -C^{[t]}(\bm p), -L^{[t]}(\bm p)]  \label{eqn:mo}\\
s.t. ~& \eqref{eqn:con1}, \eqref{eqn:con2}, \eqref{eqn:con3}. \nonumber
\end{align}
\end{subequations}

\subsection{Multi-Objective Routing Using Global Information}
In the following, we first solve problem $\mathsf{P}^{[t]}_{D, C, L}$ assuming global information is available. The $\varepsilon$-constraint method~\cite{chankong1983multiobjective} has been developed for general MOO problems, which optimizes one of the metric in the OF and treats the remaining metrics as constraints. For the multi-objective routing problem $\mathsf P^{[t]}_{D, C, L}$, the following $\varepsilon$-constraint problem $\mathsf P^{[t]}_D(\varepsilon_C, \varepsilon_L)$ can be formulated
\begin{subequations}\label{eqn:epsilon}
	\begin{align}
	\mathsf P^{[t]}_{D}(\varepsilon_C, \varepsilon_L): \quad \min_{\bm p } ~& D (\bm p) \\
	s.t. ~&C^{[t]}(\bm p) > \varepsilon_C  \label{eqn:capcity_c}\\
	&L^{[t]}(\bm p) > \varepsilon_L \label{eqn:lifetime_c}\\
   	& \eqref{eqn:con1}, \eqref{eqn:con2}, \eqref{eqn:con3}, \nonumber
	\end{align}
\end{subequations}
where we consider E2E delay minimization as the objective while treat path capacity and lifetime as constraints because the resulting problem can be solved effortlessly by modifying a standard shortest path algorithm as shown later.

In general, a \emph{weak Pareto-optimal} solution can be obtained by solving the $\varepsilon$-constraint problem~\cite{chankong1983multiobjective}. Then, by varying the values of $\varepsilon$, multiple weak Pareto-optimal solutions can be found. In particular, according to \cite[Theorem 4.2]{chankong1983multiobjective}, we can obtain the following proposition.

\begin{proposition}
If the optimal solution of the $\varepsilon$-constraint problem $\mathsf P^{[t]}_{D}(\varepsilon_C, \varepsilon_L)$ is unique, it is also a Pareto-optimal solution of $\mathsf P^{[t]}_{D, C, L}$.
\end{proposition}

As for the AANET routing problem, since the link delay $D^{[t]}(i,j)$ is a strictly increasing function of distance $d^{[t]}(i,j)$, which can be regarded as a random variable due to the uncertainty of flight position in 3D space, we can safely assume that ${\rm Pr}\big(D^{[t]}(\bm p_1) = D^{[t]}(\bm p_2)\big) = 0$ for arbitrary $\bm p_1$ and $\bm p_2$ that satisfies $\bm p_1 \neq \bm p_2$. Therefore, the optimal solution of $\mathsf P^{[t]}_D(\varepsilon_C, \varepsilon_L)$ is unique. Then, by appropriately changing the value of $\varepsilon_C, \varepsilon_L$ and solving problem~$\mathsf P^{[t]}_D(\varepsilon_C, \varepsilon_L)$, a series of Pareto-optimal solutions can be found.

Considering~\eqref{eqn:capacity}, the minimum capacity constraint \eqref{eqn:capcity_c} is equivalent to 
$
\min\limits_{n=1,\cdots, N-1}C^{[t]} (p_n,$ $p_{n+1}) > \varepsilon_C
$, which is further equivalent to $C^{[t]} (p_n, p_{n+1}) > \varepsilon_C, \forall n = 1, \cdots, N-1
$. Similarly, considering~\eqref{eqn:lifetime}, the minimum lifetime constraint~\eqref{eqn:lifetime_c} is equivalent to $L^{[t]} (p_n, p_{n+1}) > \varepsilon_L, \forall n = 1, \cdots, N-1$. Therefore, problem $\mathsf P^{[t]}_D(\varepsilon_C, \varepsilon_L)$ can be solved effortlessly by first deleting  all the links that have a capacity no higher than $\varepsilon_C$ or have a lifetime no longer than $\varepsilon_L$, and then applying any shortest path search algorithm. In Algorithm~\ref{alg:epsilon}, we modified the Floyd-Warshall algorithm for solving problem $\mathsf P^{[t]}_D(\varepsilon_C, \varepsilon_L)$ as an illustration.
\begin{algorithm}
	\caption{Modified the Floyd-Warshall algorithm for solving $\mathsf P^{[t]}_D(\varepsilon_C, \varepsilon_L)$}
	\label{alg:epsilon}
	\begin{algorithmic}[1]
		\Procedure{ComputeMinimumDelay}{}
		\State Initialize ${\sf dist}[i,j] = \sf inf$, ${\sf next}[i,j]=\sf null$, $\forall i, j\in \mathcal{N}^{[t]}$
		\For{each link $i\to j$} 
		\If{$C^{[t]}(i,j) \geq \varepsilon_C$ and $L^{[t]}(i,j) \geq \varepsilon_L$}
		\State ${\sf dist}[i,j] \leftarrow D^{[t]}(i,j)$
		\EndIf
		\EndFor
		\For{each node $i$}
		\State ${\sf dist}[i,i] \leftarrow 0$
		\EndFor
		\For{$k\in \mathcal{N}^{[t]} $, $i\in \mathcal{N}^{[t]} $, and $j\in \mathcal{N}^{[t]} $}
		\If{${\sf dist}[i,j] > {\sf dist}[i,k] + {\sf dist}[k, j]$}
		\State ${\sf dist}[i,j] \leftarrow {\sf dist}[i, k] + {\sf dist}[k, j]$
		\State ${\sf next}[i,j] \leftarrow {\sf next}[i,k]$
		\EndIf
		\EndFor
		\EndProcedure
		\Procedure{FindPath}{$i_{\sf s}$, $i_{\sf d}$}
		\If{${\sf next}[i,j] = {\sf null}$}
		\State \Return $p^*= \sf null$
		\EndIf 
		\State $p_1= i_{\sf s}$, $n=1$
		\While{$p_n\neq i_{\sf d}$}
		\State $p_{n+1} \leftarrow {\sf next}[p_n, i_{\sf d}]$ 
		\State $n \leftarrow n+1$ 
		\EndWhile
		\State \Return $\bm p^* = (p_1, p_2, \cdots, p_n)$
		\EndProcedure{}
	\end{algorithmic}
\end{algorithm}

\begin{figure*}
	\centering
	\includegraphics[width=0.75\textwidth]{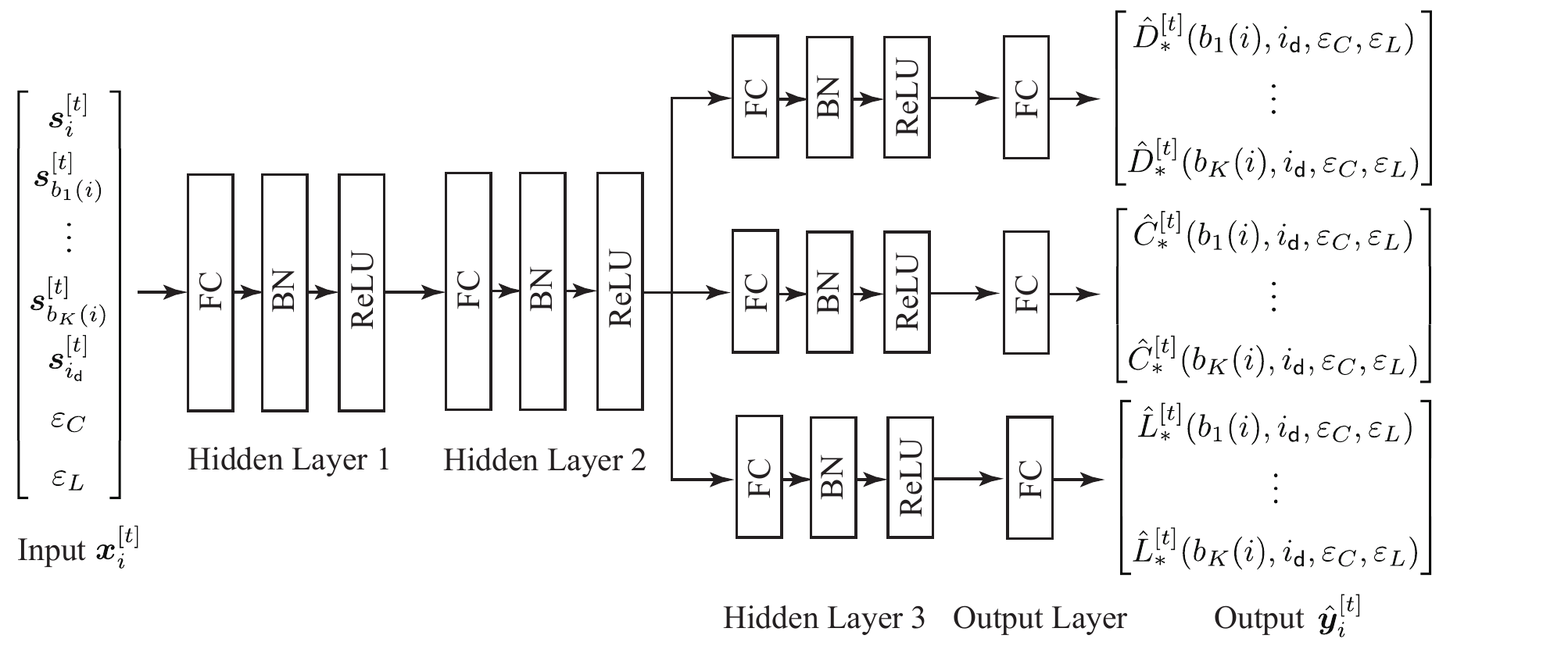}
	\caption{DNN architecture for multi-objective routing. For notational simplicity, we omit subscripts $\varepsilon_C$ and $\varepsilon_L$ in both $\bm x^{[t]}_{i, \varepsilon_C, \varepsilon_L}$ and $\hat{\bm y}^{[t]}_{i, \varepsilon_C, \varepsilon_L}$.}
	\label{fig:dnn-mo}
\end{figure*} 
In Algorithm~\ref{alg:mo}, we show how to choose the value of $\varepsilon_C$ and $\varepsilon_L$ to find all the Pareto-optimal solutions of $\mathsf P^{[t]}_{D, C, L}$. Moreover, we have the following proposition regarding the Pareto optimality of Algorithm~\ref{alg:mo}.

\begin{algorithm}
	\caption{POMOR: Find all the Pareto-optimal solutions of~$\mathsf P^{[t]}_{D, R, L}$}
	\label{alg:mo}
	\begin{algorithmic}[1]
		\State Initialize $\varepsilon_L \leftarrow 0$, Pareto set $\mathcal{P} \leftarrow \emptyset$ 
		\While{$\varepsilon_L \neq {\sf null}$}
		\State Initialize $\varepsilon_C \leftarrow 0$, $\mathcal{L} \leftarrow \emptyset$
		\While{$\varepsilon_C\neq {\sf null}$}
		\parState{Obtain the optimal solution $\bm p^*_{i_{\sf s}}$ of problem $\mathsf P^{[t]}_D(\varepsilon_C, \varepsilon_L)$ using any shortest path algorithm.}
		\If{$\bm p^*_{i_{\sf s}}={\sf null}$}
		\State $\varepsilon_C \leftarrow {\sf null}$, $\varepsilon_L \leftarrow {\sf null}$
		\Else
		\State $\varepsilon_C = D^{[t]}(\bm p^*_{i_{\sf s}})$
		\parState{Add $\bm p^*_{i_{\sf s}}$ and $L^{[t]}(\bm p^*_{i_{\sf s}})$ into sets $\mathcal{P}$ and $\mathcal{L}$, respectively.}
		\EndIf 
		\EndWhile
		\State $\varepsilon_L \leftarrow \min{\mathcal{L}} $
		\EndWhile
		\Ensure The set of Pareto-optimal solutions $\mathcal{P}$
	\end{algorithmic}
\end{algorithm}

\begin{proposition} \label{prop:2}
	All the solutions found by Algorithm~\ref{alg:mo} are the Pareto-optimal solutions of ~$\mathsf P^{[t]}_{D, C, L}$ and all the Pareto-optimal solutions of~$\mathsf P^{[t]}_{D, C, L}$ can be found by Algorithm~\ref{alg:mo}. 
\end{proposition}
\begin{IEEEproof}
	See Appendix.
\end{IEEEproof}

The complexity of Algorithm~\ref{alg:mo} is upper bounded by $\mathcal{O}(N_{\sf p}^2 N_{\varepsilon})$, where $N_{\sf p}$ is the total number of Pareto-optimal solutions of problem $\mathsf{P}_{D,C,L}^{[t]}$and $N_\varepsilon$ is the complexity of solving $\mathsf{P}^{[t]}_D(\varepsilon_C, \varepsilon_L)$. Specifically, when using the Floyd-Warshall algorithm, we have $N_\varepsilon = \mathcal{O}(|\mathcal N^{[t]}|^3) $. When using Dijkstra's algorithm, we have $N_\varepsilon = \mathcal{O}(N_{\sf e} + |\mathcal N^{[t]}| \log |\mathcal N^{[t]}|)$, where $N_{\sf e} \leq \big\vert\mathcal N^{[t]}\big\vert^2$ is the number of all possible links in the network.

\subsection{DL-Aided Multi-Objective Routing Using Local Information}
Again, global information is not available in practical AANETs. Hence,  in the following, we propose the DL-aided multi-objective routing algorithm  relying solely on local information.

Since the MOO problem $\mathsf P^{[t]}_{D, C, L}$ can be solved by solving a series of $\varepsilon$-constraint problem $\mathsf P^{[t]}_D(\varepsilon_C, \varepsilon_L)$, we propose the corresponding DL-aided routing algorithm for solving problem $\mathsf P^{[t]}_D(\varepsilon_C, \varepsilon_L)$  relying solely on local information in the following. Similar to the minimum delay routing problem $\mathsf P^{[t]}_D$, $\mathsf P^{[t]}_D(\varepsilon_C, \varepsilon_L)$ also has an optimal substructure, which can be formulated as
\begin{align}
&p_{n+1}^* = ~\arg\min_{j\in\mathcal{B}_{p_n}^{[t]}} \Big\{  D^{[t]}(p_n, j) +   D_*^{[t]}(j, i_{\sf d}, \varepsilon_C, \varepsilon_L) \nonumber \\
&+ I_{\sf inf} \left[C^{[t]}(p_n, j) \leq \varepsilon_C~\text{or}~C_*^{[t]}(j, i_{\sf d},\varepsilon_C, \varepsilon_L) \leq \varepsilon_C \right]  &\nonumber \\
&+ I_{\sf inf}\big[ L^{[t]}(p_n, j) \leq \varepsilon_L ~\text{or}~L_*^{[t]}(j, i_{\sf d}, \varepsilon_C, \varepsilon_L)\leq \varepsilon_L \big] \Big\}&
\label{eqn:next-hop-multi}
\end{align} 
where $I_{\sf inf}(X) = \sf Inf$ if $X$ is true and $I_{\sf inf}(X) = 0$ otherwise,  $D_*^{[t]}(j, i_{\sf d}, \varepsilon_C, \varepsilon_L)$, $C_*^{[t]}(j, i_{\sf d}, \varepsilon_C, \varepsilon_L)$ and $L_*^{[t]}(j, i_{\sf d}, \varepsilon_C, \varepsilon_L)$ are defined as follows. When there exists a path spanning from node $j$ to node $i_{\sf d}$ satisfying the minimum capacity constraint $\varepsilon_C$ and minimum lifetime constraint $\varepsilon_L$, $D_*^{[t]}(j, i_{\sf d}, \varepsilon_C, \varepsilon_L)$ and $C_*^{[t]}(j, i_{\sf d}, \varepsilon_C, \varepsilon_L)$ represent the delay, capacity, and lifetime achieved by the minimum-delay path satisfying the capacity and lifetime constraints. In particular, when there is no path from $j$ to $i_{\sf d}$ that satisfies the capacity and lifetime constraint, we define $D_*^{[t]}(j, i_{\sf d}, \varepsilon_C, \varepsilon_L)$, $C_*^{[t]}(j, i_{\sf d}, \varepsilon_C, \varepsilon_L)$, and $L_*^{[t]}(j, i_{\sf d}, \varepsilon_C, \varepsilon_L) $ as the delay, capacity, and lifetime achieved by the path from $j$ to $i_{\sf d}$ whose capacity and lifetime are closest to $\varepsilon_C$ and $\varepsilon_L$, respectively.

Since  $[D^{[t]}(p_n, j), C^{[t]}(p_n, j), L^{[t]}(p_n, j)]$ can be readily measured by the forwarding node $p_n$, the optimal next hop can be determined locally based on~\eqref{eqn:next-hop-multi}, once the value of $[D_*^{[t]}(j, i_{\sf d}, \varepsilon_C, \varepsilon_L), C_*^{[t]}(j, i_{\sf d}, \varepsilon_C, \varepsilon_L),$ $L_*^{[t]}(j, i_{\sf d}, \varepsilon_C, \varepsilon_L)]$ of each next-hop candidate $j\in \mathcal{C}_{i}$ is available at $p_n$. Again, we use a DNN for estimating $\big[D_*^{[t]}(j, i_{\sf d}, \varepsilon_C, \varepsilon_L), $  $ C_*^{[t]}(j, i_{\sf d}, \varepsilon_C, \varepsilon_L), L_*^{[t]}(j, i_{\sf d}, \varepsilon_C, \varepsilon_L)\big]$. The structure of the DNN is shown in Fig.~\ref{fig:dnn-mo}. 
Specifically, the input is 
\begin{equation}
\bm x_{i, \varepsilon_C, \varepsilon_L}^{[t]} = \left[\tilde{\bm s}_{i}^{[t]}, \tilde{\bm s}_{b_1(i)}^{[t]}, \cdots, \tilde{\bm s}_{b_K(i)}^{[t]}, \tilde{\bm s}_{i_{\sf d}}^{[t]}, \varepsilon_C, \varepsilon_L\right] \in \mathbb{R}^{3(K+2) + 2},
\end{equation}
where $\tilde {\bm s}_i^{[t]} = \big[\theta_i^{[t]}, \varphi_i^{[t]}, h_i^{[t]}, v_i^{[t]}, \delta_i^{[t]}\big]$.
Compared with~\eqref{eqn:x1}, the speed $v_i^{[t]}$ and heading $\delta_i^{[t]}$ of the airplane are further added into the feature, because the path lifetime depends on them. Moreover, $\varepsilon_C$ and $\varepsilon_L$ are also added into the feature to reflect the path capacity and lifetime requirements. The desired output of the DNN is
\begin{align}
&\bm y^{[t]}_{i,\varepsilon_C,\varepsilon_L}\!\!\! = \Big[ D^{[t]}_*(b_1(i), i_{\sf d}, \varepsilon_C, \varepsilon_L), \cdots, D_*^{[t]}(b_K(i), i_{\sf d},  \varepsilon_C, \varepsilon_L), \nonumber\\
&\quad~ C^{[t]}_*(b_1(i), i_{\sf d}, \varepsilon_C, \varepsilon_L), \cdots, C_*^{[t]}(b_K(i), i_{\sf d},  \varepsilon_C, \varepsilon_L), \nonumber\\
&\quad~ L^{[t]}_*(b_1(i), i_{\sf d}, \varepsilon_C, \varepsilon_L), \cdots, L_*^{[t]}(b_K(i), i_{\sf d},  \varepsilon_C, \varepsilon_L),\Big].
\end{align}

Again, we use $\bm \theta$ to denote the paremeters of the DNN. Then, the actual output of the DNN can be expressed as
\begin{align}
&\hat{\bm y}_{i, \varepsilon_C, \varepsilon_L}^{[t]}\!\! = \bm f(\bm x_{i, \varepsilon_C, \varepsilon_L}^{[t]}; \bm \theta) \nonumber\\
&=\Big[ \hat D^{[t]}_*(b_1(i), i_{\sf d}, \varepsilon_C, \varepsilon_L), \cdots, \hat D_*^{[t]}(b_K(i), i_{\sf d},  \varepsilon_C, \varepsilon_L), \nonumber\\
&\hphantom{=\Big[}~ \hat C^{[t]}_*(b_1(i), i_{\sf d}, \varepsilon_C, \varepsilon_L), \cdots, \hat C_*^{[t]}(b_K(i), i_{\sf d},  \varepsilon_C, \varepsilon_L), \nonumber\\
&\hphantom{=\Big[}~ \hat L^{[t]}_*(b_1(i), i_{\sf d}, \varepsilon_C, \varepsilon_L), \cdots, \hat L_*^{[t]}(b_K(i), i_{\sf d},  \varepsilon_C, \varepsilon_L)\Big].
\end{align}

\begin{algorithm}
	\caption{Generate training samples}
	\label{alg:train}
	\begin{algorithmic}[1]
		\State Initialize $\mathcal{E}_L = \{\varepsilon_L^{(0)} + \alpha \Delta_L\}_{\alpha =1, \cdots, A}$ and $\mathcal{E}_C = \{\varepsilon_C^{(0)} + \beta \Delta_C\}_{\beta=1, \cdots, B}$
		\For {$t\in \mathcal{T_{\sf train}}$}
			\State $D'[i_{\sf s}] \leftarrow \sf Inf$, $C'[i_{\sf s}] \leftarrow 0$, $L'[i_{\sf s}] \leftarrow 0$ for all $i_{\sf s} \in \mathcal{N}^{[t]}$
			\For{$\alpha = 0, 1, \cdots, A$} 
				\parState{$D''[i_{\sf s}] \leftarrow D'[i_{\sf s}]$, $C''[i_{\sf s}] \leftarrow C'[i_{\sf s}]$, $L''[i_{\sf s}] \leftarrow L'[i_{\sf s}]$ for all $i_{\sf s} \in \mathcal{N}^{[t]}$.}
				\For{$\beta = 0, 1 \cdots, B$}  
				\parState{Find the optimal solution $\bm p^*_{i_{\sf s}}$ of $\mathsf P^{[t]}_D( \varepsilon_L^{(0)} + \alpha \Delta_L, \varepsilon_C^{(0)} + \beta \Delta_C)$ for all $i_{\sf s}\in \mathcal{N}^{[t]}$ using Algorithm~\ref{alg:epsilon}.}
					\For{$i_{\sf s} \in \mathcal{N}^{[t]}$}
					\If{$\bm p^*_{i_{\sf s}} \neq \sf null$}
						\State\addtolength{\jot}{-0.3em}$\begin{aligned}[t] 
						&D_*^{[t]}(i_{\sf s}, i_{\sf d}, \varepsilon_C, \varepsilon_L) \leftarrow D(\bm p^*_{i_{\sf s}}), \\
						&C_*^{[t]}(i_{\sf s}, i_{\sf d} 
						\varepsilon_C, \varepsilon_L) \leftarrow C(\bm p^*_{i_{\sf s}}), \\
						&L_*^{[t]}(i_{\sf s}, i_{\sf d}, \varepsilon_C, \varepsilon_L) \leftarrow L(\bm p^*_{i_{\sf s}}).
						\end{aligned}$
						\parState{$D''[i_{\sf s}] \leftarrow D(\bm p^*_{i_{\sf s}})$, $C''[i_{\sf s}] \leftarrow C(\bm p^*_{i_{\sf s}})$, $L''[i_{\sf s}] \leftarrow L(\bm p^*_{i_{\sf s}})$.}
						\If{$\beta = 0$}
							\parState{$D'[i_{\sf s}] \leftarrow D(\bm p^*_{i_{\sf s}})$, $C'[i_{\sf s}] \leftarrow C(\bm p^*_{i_{\sf s}})$, $L'[i_{\sf s}] \leftarrow L(\bm p^*_{i_{\sf s}})$.}
						\EndIf
					\Else
						\State$\begin{aligned}[t]
						&D_*^{[t]}(i_{\sf s}, i_{\sf d}, \varepsilon_C, \varepsilon_L) \leftarrow D''[i_{\sf s}], \\
						&C_*^{[t]}(i_{\sf s}, i_{\sf d}, \varepsilon_C, \varepsilon_L) \leftarrow C''[i_{\sf s}], \\
						&L_*^{[t]}(i_{\sf s}, i_{\sf d}, \varepsilon_C, \varepsilon_L) \leftarrow L''[i_{\sf s}].
						\end{aligned}$
					\EndIf
					\If{$D_*^{[t]}(i_{\sf s}, i_{\sf d}, \varepsilon_C, \varepsilon_L) \neq \sf Inf$}
						\parState{Save $\big(\bm x_{i_{\sf s}}^{[t]}$, $\bm y_{i_{\sf s}, \varepsilon_C, \varepsilon_L}^{[t]}\big)$ as a training sample.}
					\EndIf
					\EndFor
				\EndFor
			\EndFor
		\EndFor
	\end{algorithmic}
\end{algorithm}

Algorithm~\ref{alg:train} provides the details of generating training samples from historical flight data. The parameter $\bm \theta$ can be learned by minimizing the square error between the actual output and the desired output of the DNN, which is formulated  as 
\begin{multline}
\min_{\bm \theta}~ \frac{1}{N_{\sf total}}\times \\ \sum_{t\in \mathcal{T}}\sum_{i\in \mathcal{N}^{[t]}} \sum_{\varepsilon_C \in \mathcal{E}_C} \sum_{\varepsilon_L\in \mathcal{E}_L}\!\!\!\left\Vert \bm y_{i, \varepsilon_C, \varepsilon_L}^{[t]} - \bm f(\bm x_{i, \varepsilon_C, \varepsilon_L}^{[t]}; \bm \theta)  \right\Vert^2.
\end{multline}
where $\mathcal{E}_C$ and $\mathcal{E}_L$ are sets of training samples for $\varepsilon_C$ and $\varepsilon_L$, respectively.

\begin{figure*}
	\begin{flalign}
	\tilde D_*^{[t]}(j_1, i_{\sf d}, \varepsilon_C, &\varepsilon_L) =  \left[1 - I_{\sf d}(j_1)\right] \min_{j_2\in\mathcal{C}_{j_1}^{[t]}} \bigg\{  D(j_1, j_2)  +\lambda\left[\varepsilon_C - C^{[t]}(j_1, j_2)\right]^+ + \lambda\left[\varepsilon_L - L^{[t]}(j_1, j_2)\right]^+ &\nonumber \\
	& + \left[1 - I_{\sf d}(j_2)\right]  \! \cdot \!\! \left[\hat D_*^{[t]}(j_2, i_{\sf d}, \varepsilon_C, \varepsilon_L) + \lambda\left[\varepsilon_C - \hat C_*^{[t]}(j_2, i_{\sf d},\varepsilon_C, \varepsilon_L)\right]^+   +  \lambda\left[\varepsilon_L -\hat L_*^{[t]}(j_2, i_{\sf d},\varepsilon_C, \varepsilon_L)\right]^+ \right]\bigg\}  \!\!\!\!\!\!\label{eqn:1st-2}&
	\end{flalign}
	\begin{flalign}
	\tilde D_*^{[t]}(m_k, i_{\sf d}, \varepsilon_C, \varepsilon_C)  \leftarrow  \!\!\min\limits_{j_2 \in \bar{\mathcal{C}}_{m_k}^{[t]}   }  \Big\{& D^{[t]} (m_k, j_2)  + \lambda\left[\varepsilon_C - C^{[t]}(m_k, j_2)\right]^+ + \lambda\left[\varepsilon_L - L^{[t]}(m_k, j_2)\right]^+ \nonumber\\
	& +  \left[1- I_{\sf d}(j_2)\right]  \tilde D_*^{[t]}(j_2, i_{\sf d}, \varepsilon_C, \varepsilon_L)  \Big\}    \label{eqn:2nd-2}&  
	\end{flalign}
	\hrule
\end{figure*}
Again, we can use Adam for training the DNN. After sufficient training, the DNN can be copied to each airplane for supporting the online routing decisions. Similar to Algorithm~\ref{alg:single_DL}, the DL-aided routing algorithm conceived for solving $\mathsf P^{[t]}_{D}(\varepsilon_C, \varepsilon_L)$ can be obtained by exploiting the optimal substructure~\eqref{eqn:next-hop-multi} instead, i.e., replacing~\eqref{eqn:1st} by \eqref{eqn:1st-2}, replacing~\eqref{eqn:2nd} by~\eqref{eqn:2nd-2}, and finally replacing~\eqref{eqn:pn} by
\begin{align}
& p_{n+1} = \arg\min\limits_{j \in \mathcal{C}_{p_n}^{[t]}}  \Big\{ D^{[t]} (p_n, j) + \left[1 - I_{\sf d}(j)\right] \tilde D_*^{[t]}(j, i_{\sf d},\varepsilon_C, \varepsilon_L)  \nonumber\\ 
&+ \lambda\left[\varepsilon_L - L^{[t]}(p_n, j)\right]^+ +  \lambda\left[\varepsilon_C - C^{[t]}(p_n, j)\right]^+  \Big\}, \label{eqn:pn-2}
\end{align} 
where $\lambda$ is the penalty coefficient introduced for penalizing the violation of the capacity or lifetime constraint. In practice, $\lambda$ is set to a limited value instead of infinity used in \eqref{eqn:next-hop-multi}. This is because when $\lambda$ is excessive large, \eqref{eqn:1st-2}, \eqref{eqn:2nd-2} and \eqref{eqn:pn-2} become extremely sensitive to the estimation error.

\begin{algorithm}[!htb]
	\caption{DL-aided multi-objective routing with given $\varepsilon_C$ and $\varepsilon_L$}\
	\label{alg:multi_DL}
	\begin{algorithmic}[1]
		\Algtop{$\tt \%~Obtain~the~information~for~determining~the$ $\tt next~hop$}
		\State The forwarding node $p_n$ discovers its next-hop candidates $\mathcal{C}_{p_n}$ and measures $\{D^{[t]}(p_n, j_1), $ $C^{[t]}(p_n, j_1), L^{[t]}(p_n, j_1)\}$ for $j_1\in \mathcal{C}_{p_n}$.
		\State Each next-hop candidates $j_1 \in \mathcal{C}_{p_n}$ computes $\{\hat{D}_*^{[t]}(j_2, i_{\sf d}, \varepsilon_C, \varepsilon_L)$, $\hat{C}_*^{[t]}(j_2, i_{\sf d}, \varepsilon_C, \varepsilon_L)$, $\hat{L}_*^{[t]}(j_2, i_{\sf d}, \varepsilon_C, \varepsilon_L)\}$ using the DNN,  measures $\{D^{[t]}(j_1, j_2), C^{[t]}(j_1, j_2), L^{[t]}(j_1, j_2)\}$ for $j_2\in\mathcal{C}_{j_1}$, and feeds all the information back to $p_n$.
		\Algphase{$\tt\%~Computations~at~the~forwarding~node~for$ $\tt~ determining~the~next~hop$}
		\State Initialize $\tilde D_*^{[t]}(j_2, i_{\sf d}, \varepsilon_C, \varepsilon_L) \leftarrow \hat D_*^{[t]}(j_2, i_{\sf d}, \varepsilon_C, \varepsilon_L)$ for each $j_1 \in \mathcal{C}_{p_n}, j_2 \in \mathcal{C}_{j_1}$.
		\State Update $\tilde D_*^{[t]}(j_1, i_{\sf d}, \varepsilon_C, \varepsilon_L)$ using~\eqref{eqn:1st-2} for each $j_1 \in \mathcal{C}_{p_n}$
		\State Find $\mathcal{M}_{p_n}^{[t]}$ defined by~\eqref{eqn:M} and sort the element of $\mathcal{M}_{p_n}^{[t]}$ by $\tilde D_*(m_1, i_{\sf d}, \varepsilon_C, \varepsilon_L) \leq \cdots \leq \tilde D_*(m_{|\mathcal{M}_{p_n}^{[t]}|}, i_{\sf d}, \varepsilon_C, \varepsilon_L)$. 
		\For{$k = 1, \cdots, |\mathcal{M}_{p_n}^{[t]}|$}
		\State Find $\bar {\mathcal{C}}_{m_k}^{[t]}$ defined by~\eqref{eqn:barC}.
		\State Update $\tilde D_*^{[t]}(m_k, i_{\sf d}, \varepsilon_C, \varepsilon_L)$ using~\eqref{eqn:2nd-2}.
		\EndFor
		\State Transmit the packet to $p_{n+1}$ computed by~\eqref{eqn:pn-2}. 
		\State Set $n\leftarrow n+1$ and repeat steps 1 $\sim$ 9 until the packet reaches the destination.
	\end{algorithmic}
\end{algorithm}

In Algorithm~\ref{alg:multi_DL}, we summarize the procedure of the DL-aided multi-objective routing during the online decision phase.  Upon varying the values of $\varepsilon_C $ as well as $\varepsilon_L$, and then using Algorithm~\ref{alg:multi_DL} for the given $\varepsilon_C $ and $\varepsilon_L$, multiple paths can be discovered as the solutions of the multi-objective routing problem $\mathsf{P}^{[t]}_{D,C,L}$.  

\section{Simulation Results}
In this section, we evaluate and compare the performance of the proposed routing algorithms with benchmark algorithms via simulations based on real flight data.

\subsection{Real Flight Data}
To reflect different flight mobility patterns over different airspace, the flight data was collected over the North Atlantic ocean (i.e., NA scenarios) and the European continent (i.e., EU scenario) on two representative days. Specifically, the 25th of December typically has the quietest flight traffic  and the 29th of June typically has the busiest flight traffic over a year. The status of each flight within the  region considered was recorded in the format of [TS, latitude, longitude, altitude, speed, heading] for every $10$ s over the complete $24$ hours of each selected date.

\begin{figure*}[!htb]
	\vspace{-3mm}
	\centering	
	\subfigure[NA scenario, 15:00 UTC time on 25 Dec. 2017.]{
		\label{fig:na25_data} 
		\includegraphics[height=0.34\textwidth]{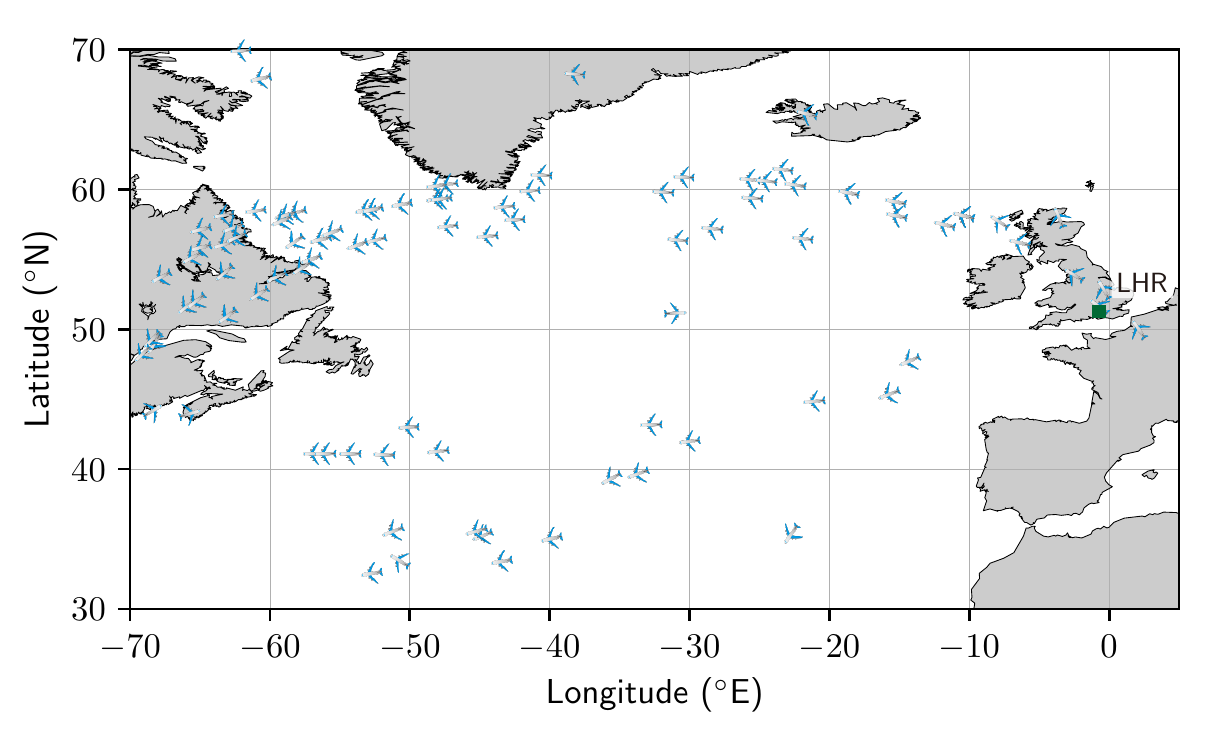}}
	\subfigure[EU scenario, 15:00 UTC time on 25 Dec. 2018.]{
		\label{fig:eu25_data} 
		\includegraphics[height=0.34\textwidth]{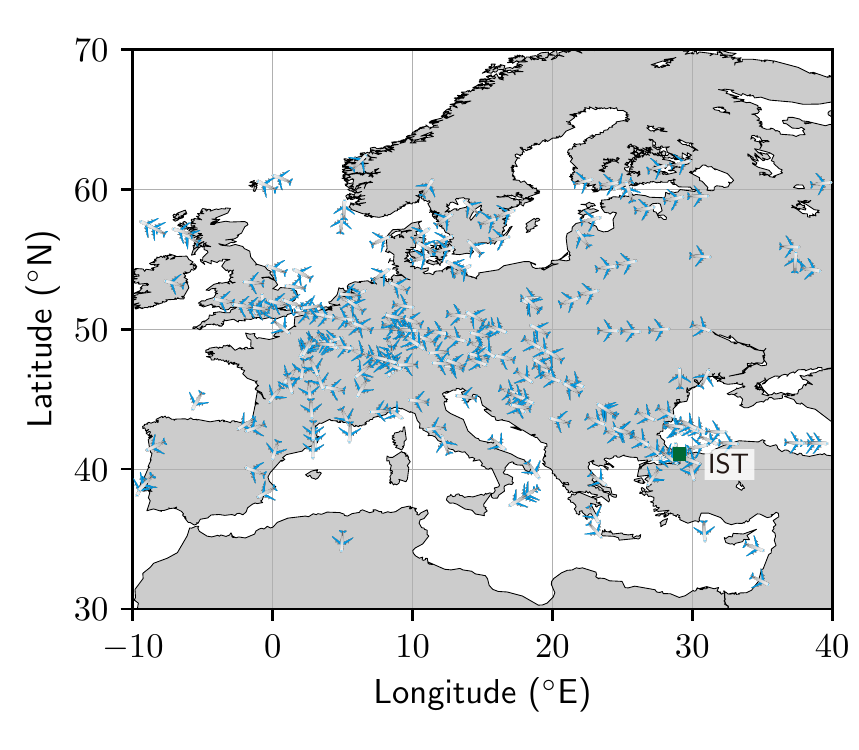}}
	\vspace{-2mm}
	\caption{Flight data over the North Atlantic ocean and the European continent.}
	\label{fig:dataset}
\end{figure*}

In Fig.~\ref{fig:dataset}, we illustrate a snapshot of the flight data at 15:00 UTC on 25 Dec. 2017 for the NA scenario and on 25 Dec. 2018 for the EU scenario. The packet destination is chosen to be the London's Heathrow Airport (LHR) and Istanbul Airport (IST) for the NA scenario and EU scenario, respectively, as labeled in Fig.~\ref{fig:na25} and Fig.~\ref{fig:eu25}.
For the NA scenario, it can be seen in Fig.~\ref{fig:na25} that strings of flights are heading towards the destination in a similar direction, while for the EU scenario shown in Fig.~\ref{fig:eu25_data}, the flight directions are quite heterogeneous.

\subsection{Simulation Settings}
Due to the limited real flight data collected, we generate more synthetic flight data based on the real flight data in order to train and test the DL-aided routing algorithms on different dataset. To reflect the mismatch between the training and testing environments, we randomly shift each flight's data along the timeline for generating the synthetic flight data for another day. By multiple realizations of the random shift based on the real flight data, multiple days of the synthetic flight data can be generated. Specially, the random shift is drawn from a Gaussian distribution with zero mean and a standard deviation of $30$ min.  We generate ten days of synthetic data based on the real flight data on Dec. 25 and Jun. 29, respectively, where six of them are used for training, three are used for validation (e.g., tuning the hyper-parameters), and one is used for testing outside the training and validation sets.

We further divided the training dataset into four time windows, each six hours, where a DNN was trained separately. In particular, for the simulations presented in the following, the DNN was trained using the training flight data within a time window of 12:00 $\sim$ 18:00 UTC. Once the DNN is well-trained, the DNN can be used for assisting the routing decisions during 12:00 $\sim$ 18:00 UTC in the testing dataset without the need of updating the DNN. 

The transmit power is $P = 30$ dBm, the antenna gain is $G_{\sf t} = G_{\sf r}=25$ dBi, and the carrier frequency is $f=14$ GHz \cite{vondra2018integration}. The transmission bandwidth is $W = 6$ MHz and the noise power is computed by $\sigma^2 = kTWF$, where $k=1.3\times 10^{-23}$, $T = 223.15$ Kelvin (i.e., $-50~{}^\circ$C at an altitude of $10$ km), $F = 4$ dB is the receiver’s noise figure~\cite{zhang2017adaptive}. The packet size is $S = 1$  KBytes. To generate the labels for training the DNN, the queuing delay is set as a constant $D_{\sf que}^{[t]}(i) = 10$ ms for each node during training. By contrast, to reflect the heterogeneous traffic load of each node during testing, the queuing delay of each node is randomly drawn from a $[1, +\infty)$-truncated Gaussian distribution with a mean value of $10$ ms and a standard deviation of $5$ ms.

All experiments are conducted on an ordinary personal computer (PC) with AMD Ryzen\texttrademark
~9 3950X CPU and
a single Nvidia Geforce RTX\texttrademark~2080Ti GPU. The DNNs are implemented using TensorFlow
1.15 \cite{tensorflow2015-whitepaper} on Windows 10.

\subsection{Hyper-Paremeter Settings for DL-aided Routing Algorithms}
Each hidden layer of the DNN in Algorithm~\ref{alg:single_DL} has $100$ neurons, as shown in Fig.~\ref{fig:dnn}. The first two hidden layers of the DNN in Algorithm~\ref{alg:multi_DL} have $300$ neurons and the third hidden layer has $100$ neurons for each stream, as shown in Fig.~\ref{fig:dnn-mo}. We use He's initialization~\cite{he2015delving} for all the hidden layers and all the output layers are initialized from the uniform distribution of $[-3\times 10^{-3}, 3\times 10^{-3}]$. The initial learning rate for the Adam optimizer is set to $10^{-3}$. The mini-batch size for gradient descent is $1000$. 
The maximal number of neighbors to be considered as the next-hop candidates is set to $10$ for Algorithm~\ref{alg:single_DL} and set to $40$ for Algorithm~\ref{alg:multi_DL}. For Algorithms~\ref{alg:train} and~\ref{alg:multi_DL}, $\varepsilon_C$ and $\varepsilon_L$ are swept over $[20, 50]$ Mbps with a step size of $2$~Mbps and $[0, 30]$ min with a step size of $5$ min, respectively, in order to discover multiple paths. The penalty coefficient is set to $\lambda = 10$.
\subsection{Performance Analysis and Comparison}
The following routing protocols are considered for comparison:

\begin{enumerate}
	\item \emph{Optimal}: The optimal path found by solving the miminal delay problem~$\mathsf P^{[t]}_D$ via the Floyd-Warshall algorithm, which relies on the global information regarding the delay of  every possible link in the network.
	\item \emph{Greedy perimeter stateless routing (GPSR)}: The routing protocol proposed in \cite{karp2000gpsr}, which is
	solely based on local geographic information. Specifically, each node forwards its received packet to the specific neighbor that is geographically closest to the destination.
	When a packet reaches a node where greedy forwarding fails, the algorithm recovers by routing around the perimeter of the region.
	\item \emph{Geographic Load Share
		Routing (GLSR)}: The routing protocol proposed in \cite{medina2011geographic}, which is based on the greedy forwarding used in GPSR and also takes the queuing delay of each next-hop candidate into consideration, when deciding about the next hop.
	\item \emph{Pareto Optimal}: The Pareto-optimal paths found by Algorithm~\ref{alg:mo}, which relies on the global information regarding the delay, capacity and lifetime of every possible link in the network.
\end{enumerate}

\begin{figure}[!htb]
	\centering	
	\subfigure[NA scenario, 15:00 $\sim$ 16:00 UTC Dec. 25, 2017.]{
		\label{fig:na25} 
		\includegraphics[width=0.43\textwidth]{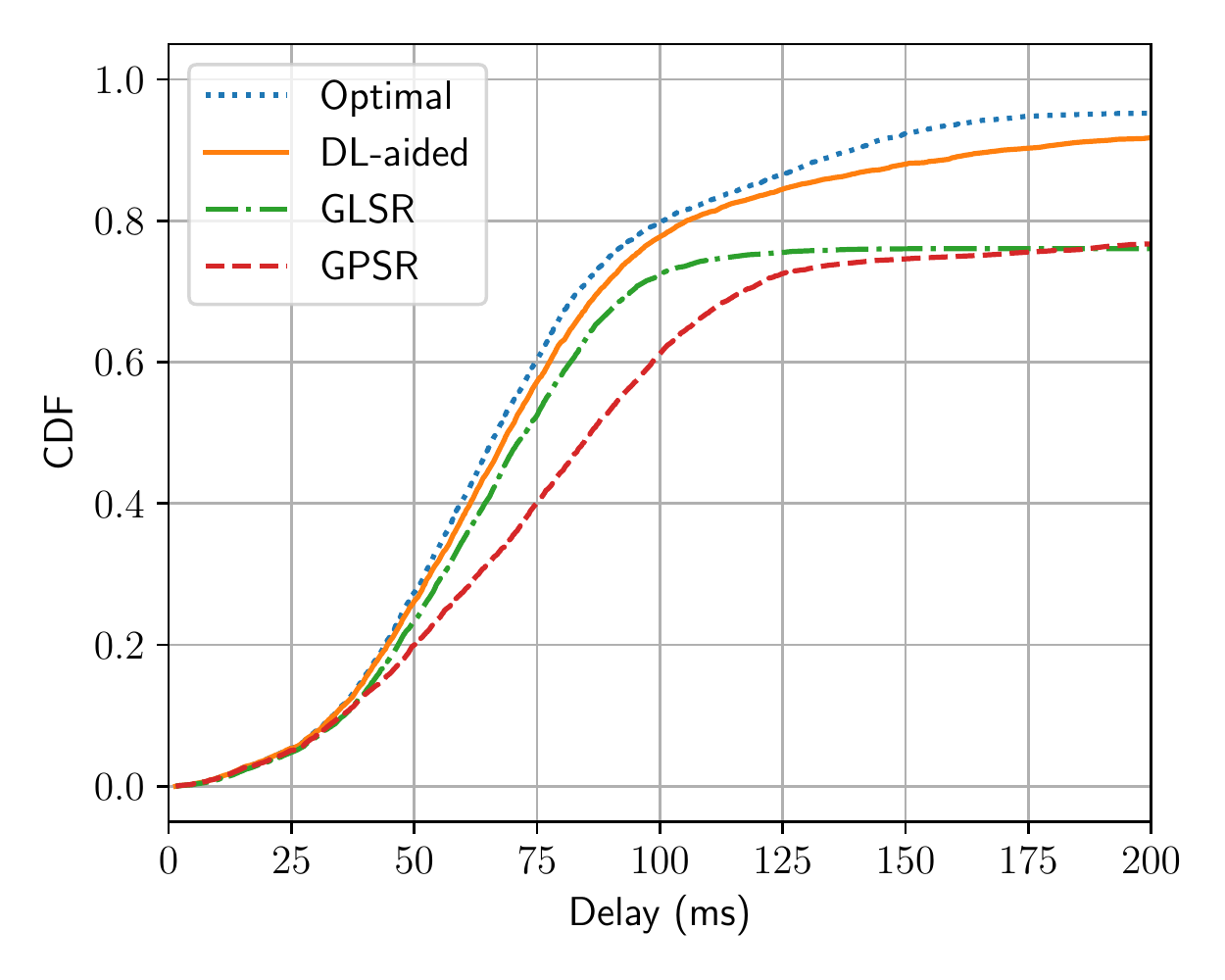}}
	\subfigure[EU scenario, 15:00 $\sim$ 16:00 UTC Dec. 25, 2018.]{
		\label{fig:eu25} 
		\includegraphics[width=0.43\textwidth]{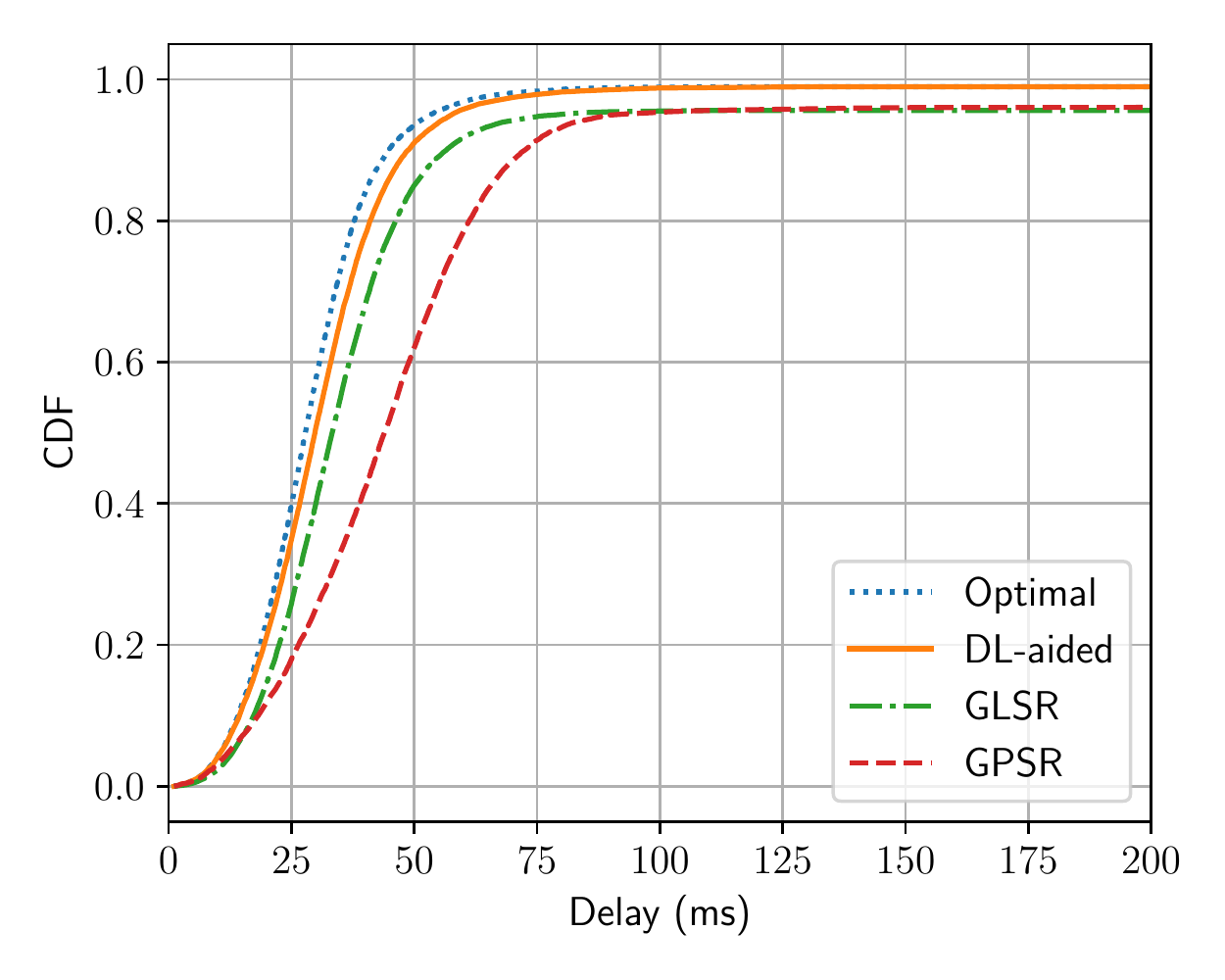}}
	\caption{E2E delay comparison for NA and EU scenarios.}
	\label{fig:delay}
\end{figure}

\begin{figure*}[!htb]
	\centering	
	\subfigure[NA scenario, 15:00 UTC Dec. 25, 2017.]{
		\label{fig:na25_path} 
		\includegraphics[height=0.35\textwidth]{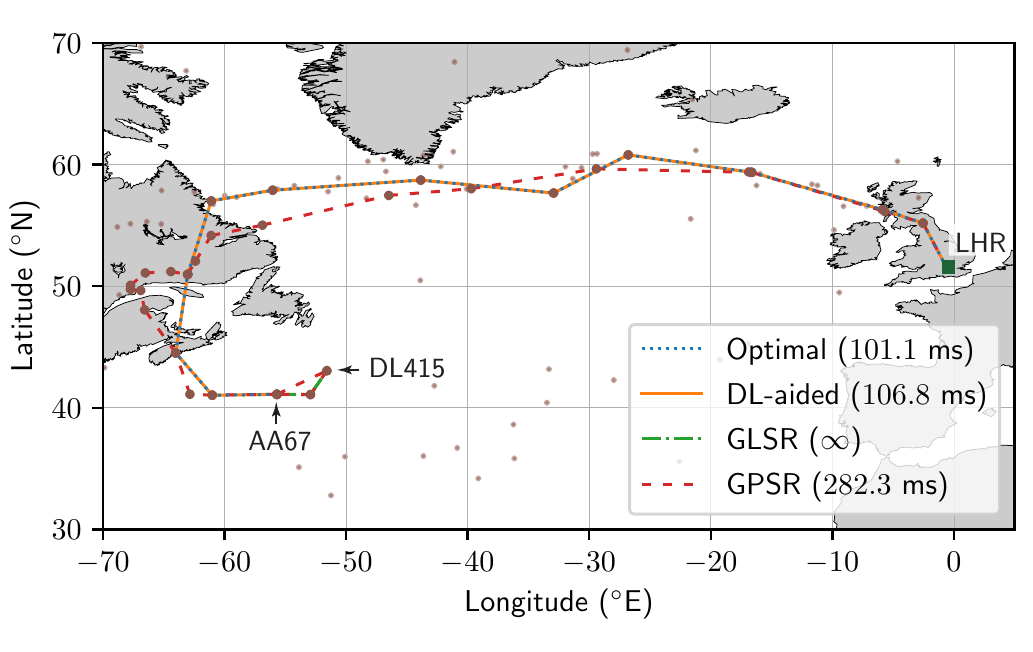}}
	\subfigure[EU scenario, 15:00 UTC Dec. 25, 2018.]{
		\label{fig:eu25_path} 
		\includegraphics[height=0.35\textwidth]{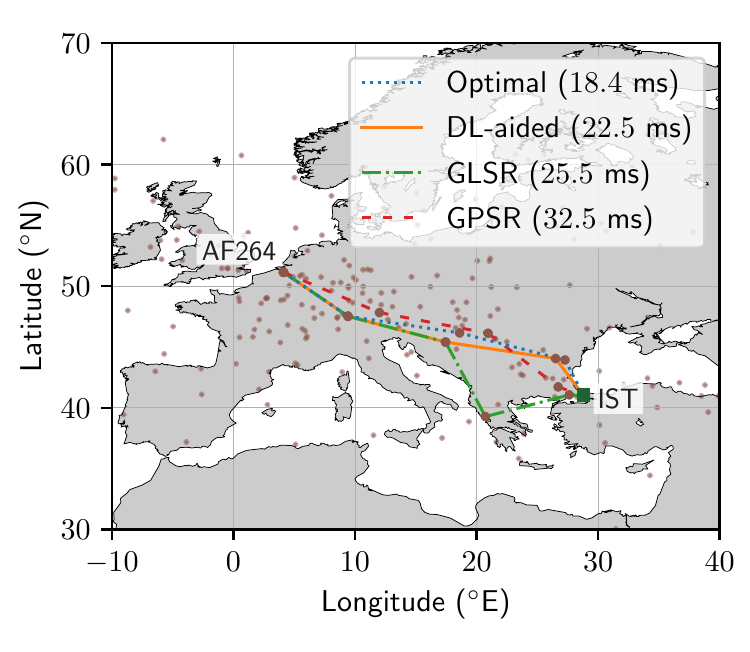}}
	\caption{Comparison of paths found by different routing algorithms. }
	\label{fig:path}
\end{figure*}

In Fig.~\ref{fig:delay}, we compare the cumulative distribution function (CDF) curves of the E2E delay of all the flights on the quietest day, which can be regarded as the worst-case scenario for AANETs. In particular, we show the results during a time window of 15:00 $\sim$ 16:00 for a clear comparison. We can see that GLSR outperforms GPSR, because the queuing delay is taken into consideration. Our proposed DL-aided routing (Algorithm~\ref{alg:single_DL}) performs close to the optimal policy, achieving lower E2E delay and higher success probability (defined as the probability of E2E delay lower than $200$ ms) than GPSR and GLSR for both the NA and EU scenario. We note that similar conclusions can be made when testing on other time windows and on other days, which are not included here due to the space limitations.

To explain how the DL-aided routing algorithm outperforms the benchmark policies, we further compare the paths and their corresponding E2E delays found by different routing algorithms in an example snapshot for a particular flight in each scenario.  In Fig.~\ref{fig:na25_path}, we show the paths found for flight AA67 to LHR. We can see that GLSR encounters a communication void, when the packet is forwarded to flight DL415, and hence fails to find a path to the packet destination. By contrast, GPSR use perimeter routing after it encounters the communication void. Although GPSR can find a path to the destination eventually, it struggles through many hops to get around the void region. Since our DL-aided routing is trained using the historical flight data and hence is embedded with the topology knowledge, it can bypass the communication void more efficiently than GPSR, and find a similar path to the optimal path. In Fig.~\ref{fig:eu25_path}, we show the paths found for flight AF264 to IST. We can see that the flight distribution over the European continent is denser, and hence it is less likely to encounter a  communication void. Consequently, all routing algorithms find paths with similar number of hops. Since GPSR does not consider the queuing delay, it achieves the highest E2E delay. In contrast, GLSR and our DL-aided routing algorithms achieve lower E2E delay. Benefited from the DNN and the feedback mechanism, each forwarding node can exploit more information using the DL-aided routing algorithm for determining the next hop, and hence DL-aided routing achieve lower E2E delay than GLSR and performs closely to the optimal path.

\begin{figure}
	\centering	
	\subfigure[Flight DL405 to LHR, 15:00 UTC Jun. 29, 2018.]{
		\label{fig:na3d} 
		\includegraphics[width=0.43\textwidth]{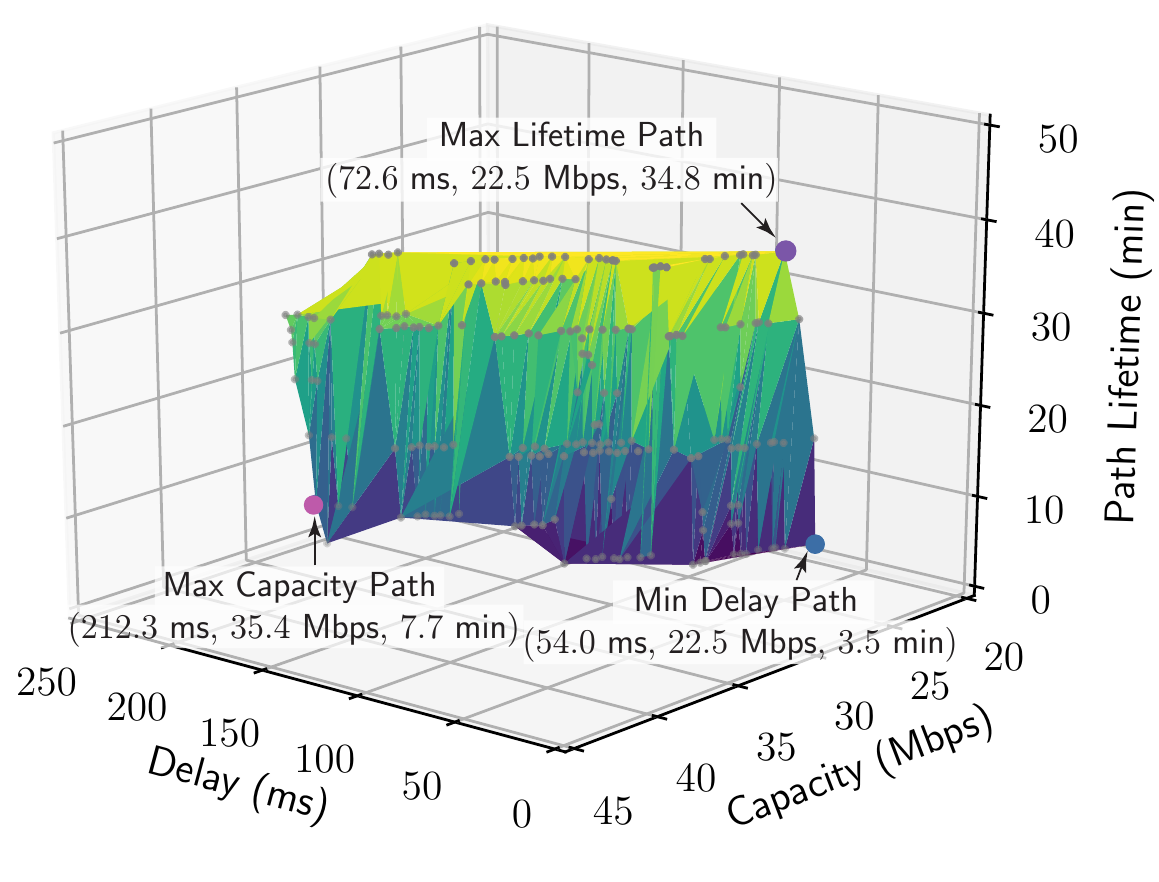}}
	\subfigure[Flight BA480 to IST, 15:00 UTC Jun. 29, 2018.]{
		\label{fig:eu3d} 
		\includegraphics[width=0.43\textwidth]{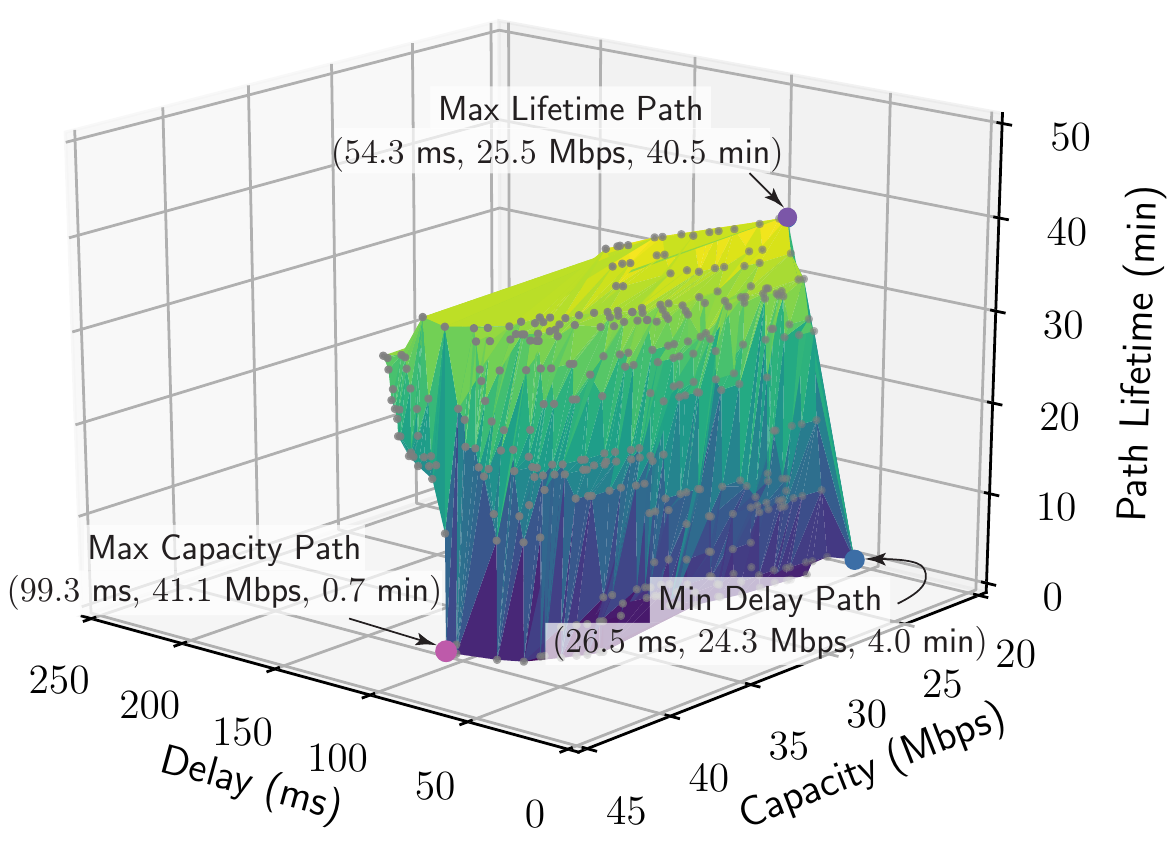}}
	\caption{3D Pareto front.}
	\label{fig:3d}
\end{figure}

\begin{figure*}
	\centering	
	\subfigure[Flight DL405 to LHR, 15:00 UTC Jun. 29, 2018.]{
		\label{fig:na3dpath} 
		\includegraphics[height=0.35\textwidth]{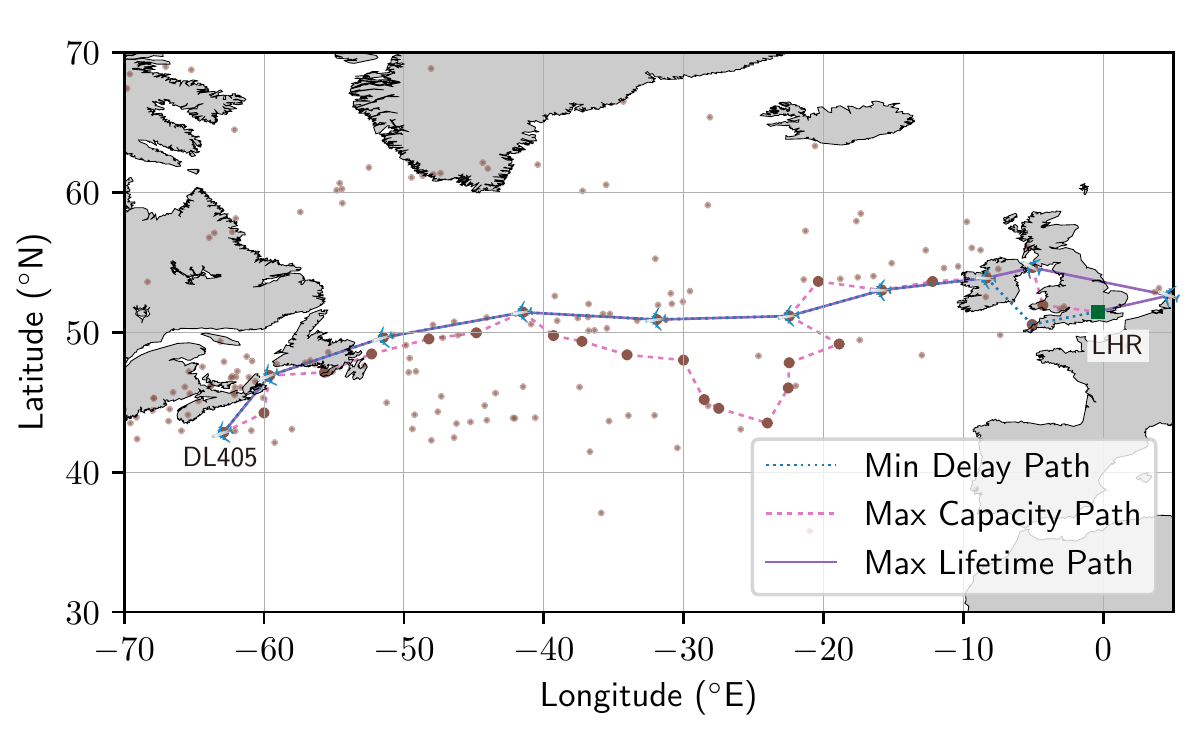}}
	\subfigure[Flight BA480 to IST, 15:00 UTC Jun. 29, 2018.]{
		\label{fig:eu3dpath} 
		\includegraphics[height=0.35\textwidth]{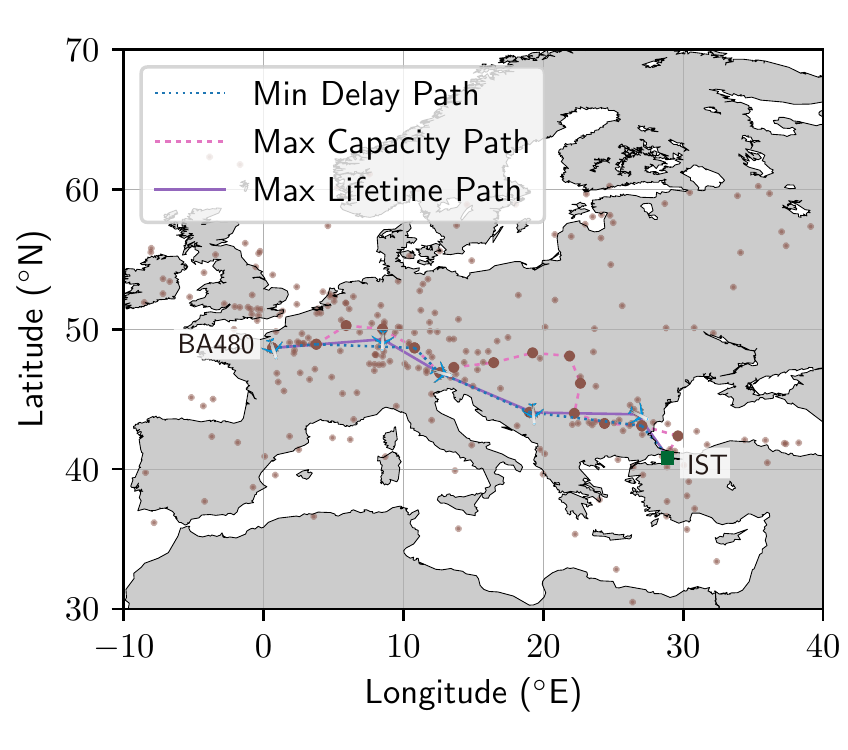}}
	\caption{Minimum-delay path, maximum-capacity path, and maximum-lifetime path. }
	\label{fig:3dpath} 
\end{figure*}

To test the routing algorithms when multi-component OF is considered, we use the flight data on the busiest day, where more paths are available between the source and destination nodes due to more airplanes in the sky. Specifically, we consider the paths from flight DL405 to LHR in the NA scenario and the paths from flight BA480 to IST in the EU scenario (both on 15:00 UTC Jun. 29, 2018) as examples for demonstrating the relationship between delay, capacity and lifetime achieved by the Pareto-optimal paths. In Fig.~\ref{fig:3d}, we plot the Pareto front of the multi-objective routing problem $\mathsf{P}^{[t]}_{D, C, L}$ found by Algorithm~\ref{alg:mo}, where each point represents a Pareto-optimal path and the surface interpolated by all the points visualizes the 3D Pareto front.  Among all the Pareto-optimal paths, we highlight the performance achieved by the minimum-delay path, the maximum-capacity path and the maximum-lifetime path. We can see that improving one metric on the Pareto front will sacrifice at least one of the other two metrics. For example, the minimum-delay path from flight DL405 to LHR has a delay of $54$ ms, a capacity of $22$ Mbps, and a lifetime of $3.5$min. To increase the capacity to its maximum value of $35.4$ Mbps, the lifetime increase to $7.7$ min simultaneously while the delay increases to $212.3$ ms. To increase the lifetime to its maximum value of $34.8$ min, the capacity remains unchanged while the delay increases to $72.6$ ms.

To further investigate the tradeoff relationship, we plot the minimum-delay path, maximum-capacity path, and maximum-lifetime path of the example snapshots in Fig.~\ref{fig:3dpath}. We can see that to achieve the maximum capacity, the number hops increases to reduce the distance of each link, which results in longer total distance from the source to the destination and higher E2E delay. To show how the maximum-lifetime path is selected, we further plot the flight direction of the nodes on the maximum-lifetime path. We can see that, to achieve the maximum path lifetime, the angle between the flight directions of adjacent nodes is small so that the lifetime of each link is maximized. Moreover, from Fig.~\ref{fig:na3dpath}, we can see that most of the maximum-lifetime path share the same nodes with the minimum-delay path. This is because the flight directions are similar in the NA scenario, where strings of flights are heading toward the destination continent.

\begin{figure}[!htb]
	\centering	
	\subfigure[Flight DL405 to LHR, 15:00 UTC Jun. 29, 2018.]{
		\label{fig:na2d} 
		\includegraphics[width=0.43\textwidth]{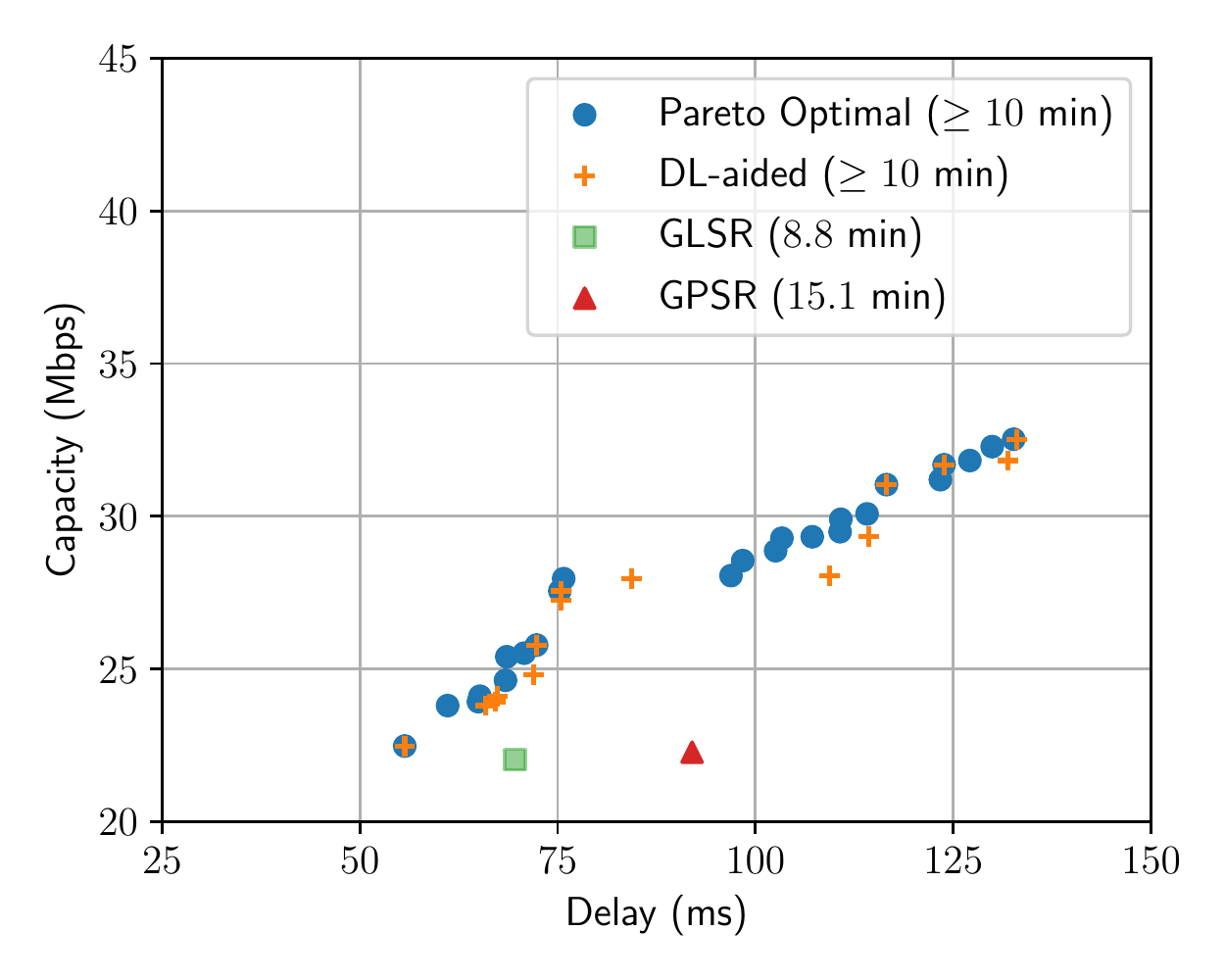}}
	\subfigure[Flight BA480 to IST, 15:00 UTC Jun. 29, 2018.]{
		\label{fig:eu2d} 
		\includegraphics[width=0.43\textwidth]{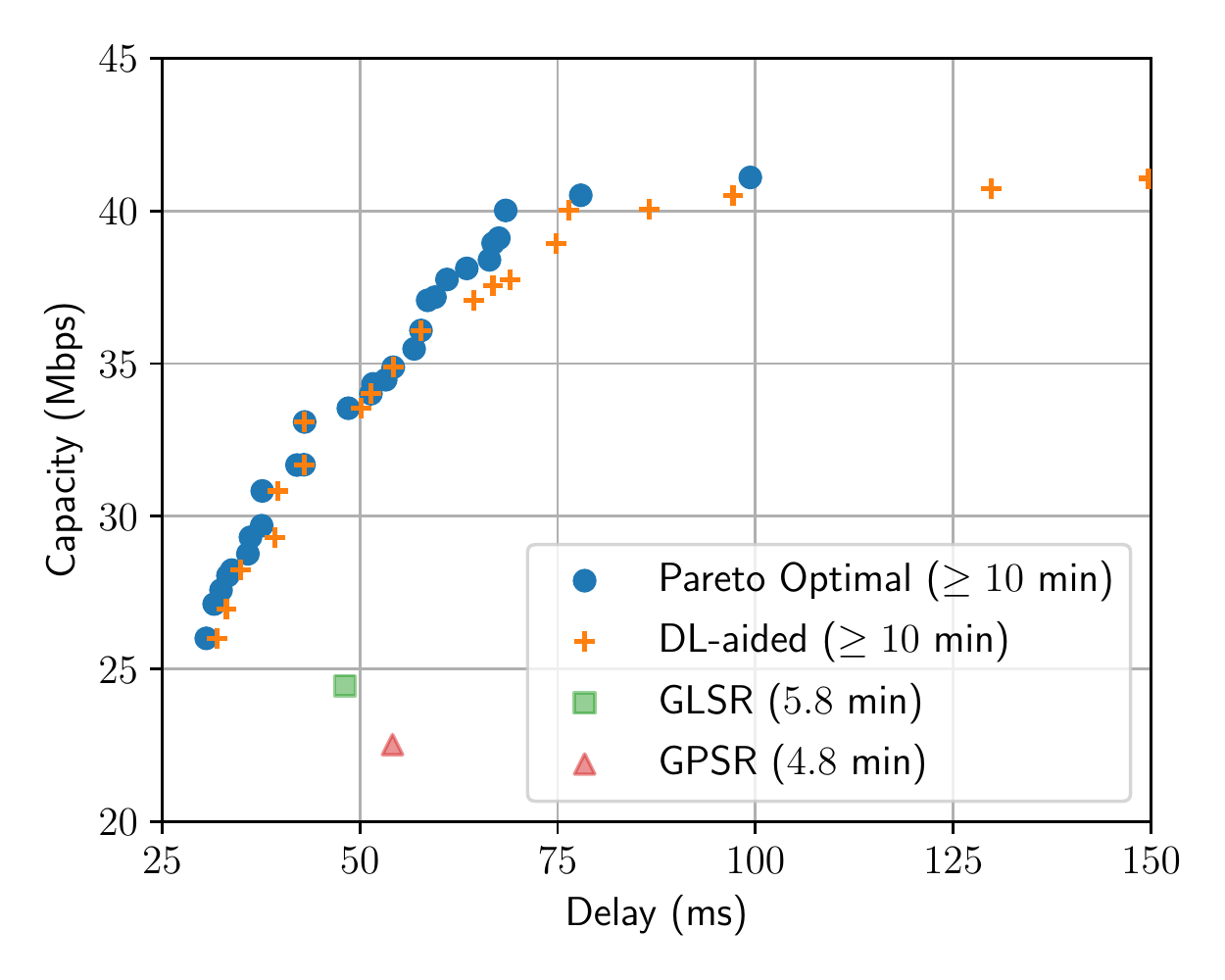}}
	\caption{Capacity and delay tradeoff. The minimum lifetime constraint for ``Pareto Optimal" and ``DL-aided" is $10$ min. }
	\label{fig:2d}
\end{figure}

Since there exist many Pareto optimal paths as shown in Fig.~\ref{fig:3d} and maximizing the path lifetime may not be so important as minimizing the delay or maximizing the capacity once the path lifetime exceeds a certain threshold, in the following, we treat the lifetime as a constraint and consider the tradeoff between capacity and delay. Specifically, the minimum lifetime constraint is set to $10$ min. We compare the capacity and delay of different routing algorithms in Fig.~\ref{fig:2d}. Since GLSR and GPSR may not guarantee the minimum lifetime constraint, we also provide the corresponding path lifetime in the legend. We can see that our DL-aided routing (Algorithm~\ref{alg:multi_DL}) can find paths having a performance approaching to the Pareto front and dominate the paths found by GLSR and GPSR. Moreover, the lifetime constraint is guaranteed simultaneously. By contrast, GLSR fails to guarantee the lifetime constraint in both the NA and EU scenarios, while GPSR can only guarantee the lifetime constraint in the NA scenario, where most of the flight directions are similar.

\begin{table}[htbp]
	\centering
	\caption{Complexity Comparison}
	\small
	\begin{tabular}{lrr}
		\toprule
		\bf Complexity Indicator & \bf SO-DNN & \bf MO-DNN \\
		\midrule
		Number of parameters to be trained & $1.5\times 10^4$ & $1.9\times 10^5$ \\
		Input feature dimensions & $36$    & $212$ \\
		Number of training samples & $9.1\times 10^4$ & $1.1\times 10^7$ \\
		Training time per iteration & $2.3$ ms & $7.9$ ms \\
		Number of iterations for training & $2\times 10^3$  & $10^4$ \\
		Total training time & $4.6$ s   & $79$ s \\
		One-shot evaluation time & $0.04$ ms & $0.09$ ms \\
		\bottomrule
	\end{tabular}%
	\label{tab:complexity}%
\end{table}%

In Table~\ref{tab:complexity}, we compare the training and testing complexity of the DNN used in single-objective routing and multi-objective routing, both in NA scenario on Jun. 29, 2018. For ease of expression, the DNN used for single-objective routing is termed as the SO-DNN, while the one used for multi-objective routing is termed as the MO-DNN. Since the MO-DNN has higher input dimension, more hidden layers and more neurons per layer for learning the relationship between the delay/capacity/lifetime and the thresholds, its number of parameters to be trained is about $13$ times higher than that of SO-DNN. Moreover, because we have to sample different thresholds for the training of the MO-DNN, its training set is about a hundred times higher than that of the SO-DNN. As a result, the number of iterations required for convergence is $5$ times higher compared to the training of SO-DNN. Further considering that the complexity per iteration of the MO-DNN's training is about $3.4$ times higher than that of SO-DNN, the total training complexity of the MO-DNN is about $17$ times higher than that of the SO-DNN, which can be done within minutes using an ordinary PC. 

For the testing part, the multi-objective routing has to evaluate the MO-DNN's output for different input thresholds in order to discover multiple paths. Therefore, the testing complexity is proportional to the number of thresholds. Fortunately, we can combine the different thresholds into a batch and evaluate the MO-DNN's output for all the thresholds in parallel. Therefore, its computation time is similar to that in SO-DNN.
Moreover, in the testing phase, only one-shot forward propagation is required for evaluating the DNN's output, which in general has significantly lower computational complexity than that of training relying on numerous iterations of back propagation. As a result, both the execution times for a one-shot evaluation for the outputs of SO-DNN and MO-DNN are less than 1 ms. 

\section{Conclusions}
In this paper, we proposed DL-aided routing policies for AANETs formed by passenger airplanes, where DNNs are invoked for learning the relationship between the local geographic information observed by the forwarding node and the information that is required for determining the optimal next hop. The DNN is trained by supervised learning based on historical flight data and it is then stored by each airplane for assisting their routing decisions during flight solely based on their local information in a distributed manner. We further extended the DL-aided routing algorithm to multi-objective scenarios, where the delay, capacity and path lifetime were jointly optimized. Our simulation results based on real flight data showed that the proposed DL-aided routing outperforms existing routing protocols in terms of their delay, capacity as well as path lifetime, and it is capable of approaching the Pareto front that is obtained using perfectly known global link information.  

It is worth noting that although AANETs can be formed in many regions where the flight-density is high enough, it may fail in certain regions where the flight-density is low. Moreover, MOO introduces extra the computational complexity compared to single-objective optimization, resulting in higher computation resource demand, as well as higher processing time and power. Future research may integrate satellites into the AANET for supporting truly global coverage and may investigate how to further reduce the complexity of MOO.

\appendix
\section{Proof of Proposition 2}
According to Proposition 1 and bearing in mind that the solution of $\mathsf P^{[t]}_D(\varepsilon_C, \varepsilon_L)$ is unique, we can see that all the solutions found by Algorithm~\ref{alg:mo} are the Pareto optimal solutions of $\mathsf P^{[t]}_{D, C, L}$. In the following, we prove the rest of Proposition~\ref{prop:2}.

For notation simplicity, we omit the superscript ``$[t]$" in what follows. Let $\bm p_{m, n}$ denote a feasible solution found by the $m$th iteration of the outer loop and the $n$th iteration of the inner loop of Algorithm~\ref{alg:mo}, and let $N_m$ denote the total number of the inner iterations of the $m$th outer iteration. Then, according to steps 9 and 11 of Algorithm~\ref{alg:mo}, $\bm p_{m, n}$ is the optimal solution of $\mathsf P_D\big(C(\bm p_{m,n-1}), \min_{n'}{L(\bm p_{m-1, n'})}\big)$, $n'\in \{1, \cdots, N_{m-1}\}$. Moreover, since $C(\bm p_{m,n}) > 0$, $\bm p_{m,n}$ is also a feasible solution of $\mathsf P_D\big(0, \min_{n'}{L(\bm p_{m-1, n'})}\big)$. We first prove the following lemma.
\begin{lemma}
There exists an $\tilde n\in\{1, \cdots N_m\}$ such that $[D(\bm p_{m, \tilde n}), -C(\bm p_{m, \tilde n})]$ dominates $[D(\bm p'), -C(\bm p')]$ for any feasible solution $\bm p'$ of $\mathsf P_D\big(0,\min_{n'}$ ${L(\bm p_{m-1, n'})}\big)$ that satisfies $\bm p' \notin \{\bm p_{m, n}\}_{n=1,\cdots, N_m}$.
\end{lemma}

\begin{IEEEproof}
	Considering step 9 of Algorithm~\ref{alg:mo} and bearing in mind that the solution of $\mathsf P^{[t]}_D(\varepsilon_C, \varepsilon_L)$ is unique, it is not hard to prove that $D(\bm p_{m, n}) < D(\bm p_{m, n + 1})$ and $C(\bm p_{m, n}) < C(\bm p_{m, n+1})$. Assume that there exists another feasible solution of $\mathsf P_D\big(0, \min_{n'}{L(\bm p_{m-1, n'})}\big)$, $\bm p' \notin \{\bm p_{m, n}\}_{n=1,\cdots,N_m}$ such that $[D(\bm p'), -C(\bm p')]$ is not dominated by $\bm p_{m, n}$ for any $n\in\{1, \cdots, N_m\}$. Further assume that there exists an $ n_0 \in \{1,\cdots, N_m-1\}$ such that $D(\bm p_{m,n_0}) < D(\bm p') < D(\bm p_{m, n_0+1})$. Since the optimal solution of $\mathsf P_D\big(\varepsilon_C, \varepsilon_L\big)$ is unique, $[D(\bm p_{m, n_0}),$ $-C(\bm p_{m, n_0})]$  and $[D(\bm p_{m, n_0+1}), -C(\bm p_{m, n_0+1})]$ are non-dominated by $[D(\bm p), C(\bm p)]$ for any feasible solution $\bm p$ of $\mathsf P_D\big(0, $ $ \min_{n'}{L(\bm p_{m-1, n'})}\big)$ according to Proposition~1. Then, we can obtain $-C(\bm p_{m, n_0}) > -C(\bm p') > -C(\bm p_{m, n_0+1})$, i.e., $C(\bm p_{m, n_0}) < C(\bm p') < C(\bm p_{m,n_0+1})$. Considering that $\bm p_{m, n_0+1}$ is the optimal solution of $\mathsf P_D\big(C(\bm p_{m, n_0}),\min_{n'}{L(\bm p_{m-1, n'})}\big)$ and $\bm p'$ is a feasible solution of $\mathsf P_D\big(C(\bm p_{m, n_0}),$ $ \min_{n'}{L(\bm p_{m-1, n'})}\big)$, we have $D(\bm p_{m, n_0+1}) \leq D(\bm p') $, which contradicts to $D(\bm p_{m,n_0}) < D(\bm p') < D(\bm p_{m, n_0+1})$. Similarly, we can readily obtain contradictions when $ D(\bm p') < D(\bm p_{m,1})$, or when $D(\bm p') > D(\bm p_{m,N_m})$. Therefore, Lemma 1 is proved.
\end{IEEEproof}

Next, we continue to prove the rest of Proposition 2. Assume that there exists another feasible solution of $\mathsf P_{D, C, L}$, i.e., $\bm p'' \notin \{ \bm p_{m, n}\}$, that is also a Pareto-optimal solution of $\mathsf P^{[t]}_{D, C, L}$. According to step 11 of Algorithm~\ref{alg:mo}, we have $\min_{n'} L(\bm p_{m-1, n'}) < \min_{n'}L(\bm p_{m, n'})$. Let $M$ denote the total number of outer iterations and assume furthermore that there exist an $m_0 \in \{1, \cdots, M\}$ such that $\min_{n'} {L(\bm p_{m_0-1,n'})}< L(\bm p'') \leq \min_{n'} {L(\bm p_{m_0,n'})}$. Since $C(\bm p'')>0$ and $L(\bm p'') > \min_{n'} {L(\bm p_{m_0-1,n'})} $, $\bm p''$ is also a feasible solution of $\mathsf P(0,  \min_{n'} {L(\bm p_{m_0-1,n'})})$. Then, according to Lemma 1, there exists an $\tilde n\in \{1,\cdots,N_{m_0}\}$ such that $[D(\bm p_{m_0, \tilde n}),$ $C(\bm p_{m_0, \tilde n})]$ dominates $[D(\bm p''),$ $C(\bm p'')]$. Moreover, since $L(\bm p'')$ $\leq \min_{n'} {L(\bm p_{m_0,n'})}$, we have $L(\bm p'') \leq L(\bm p_{m_0,\tilde n})$. Consequently, $[D(\bm p_{m_0, \tilde n}), C(\bm p_{m_0, \tilde n}),$ $L(\bm p_{m_0, \tilde n})]$ dominates $[D(\bm p''), C(\bm p''),L(\bm p'')]$, which contradicts to the assumption that $\bm p''$ is Pareto-optimal. Similarly, we can easily obtain contradictions when $L(\bm p'') \leq \min_{n'} {L(\bm p_{1,n'})}$, or when $L(\bm p'') > \min_{n'} {L(\bm p_{M,n'})} $. Therefore, all the Pareto-optimal solutions of~$\mathsf P^{[t]}_{D, C, L}$ can be found by Algorithm~\ref{alg:mo} and hence Proposition 2 is proved.
\bibliographystyle{IEEEtran}
\vspace{-2mm}
\bibliography{ref}
\end{document}